\documentclass[lettersize,journal,twoside]{IEEEtran}
\usepackage{amsmath,amsfonts}
\usepackage{algorithmic}
\usepackage{algorithm}
\usepackage{array}
\usepackage[caption=false,font=normalsize,labelfont=sf,textfont=sf]{subfig}
\usepackage{textcomp}
\usepackage{stfloats}
\usepackage{url}
\usepackage{verbatim}
\usepackage{graphicx}
\usepackage{cite}

\usepackage{enumitem}
\usepackage{multirow}
\usepackage[caption=false,font=normalsize,labelfont=sf,textfont=sf]{subfig}
\usepackage{url}
\usepackage{color}
\usepackage[usenames,dvipsnames]{xcolor}
\usepackage{comment}

\newcommand {\ie} {{\em i.e., }}
\newcommand {\eg} {{\em e.g., }}
\newcommand {\iid} { i.i.d. }
\newcommand {\etal} {{\em et al.}}
\newcommand {\beq} {\begin{equation}}
\newcommand {\eeq} {\end{equation}}
\newcommand {\bequn} {\begin{equation*}}
\newcommand {\eequn} {\end{equation*}}
\newcommand {\bear} {\begin{eqnarray}}
\newcommand {\eear} {\end{eqnarray}}
\newcommand {\bearun} {\begin{eqnarray*}}
\newcommand {\eearun} {\end{eqnarray*}}
\newcommand {\fig}[1]{Fig.~\ref{#1}}
\newcommand {\Eqref}[1]{Eq.~(\ref{#1})}

\newcommand{\add}[1]{{\color{black}{#1}}}

\DeclareMathOperator*{\argmax}{\arg\max}

\begin{document}

\title{Adaptive Edge Offloading for Image Classification Under Rate Limit
}

\author{Jiaming Qiu, Ruiqi Wang, Ayan Chakrabarti, Roch Gu\'{e}rin, Chenyang Lu \\
Dept. of Computer Science \& Engineering, Washington University in St. Louis.
\{qiujiaming,ruiqi.w,guerin,lu\}@wustl.edu, ayan.chakrabarti@gmail.com
}



\maketitle

\begin{abstract}
This paper considers a setting where embedded devices are used to acquire and classify
\add{images}. Because of limited computing capacity, embedded devices rely on a parsimonious classification model with uneven accuracy. When local classification is deemed inaccurate, devices can decide to offload 
\add{the image}
to an edge server with a more accurate but resource-intensive model. Resource constraints, \eg network bandwidth, however, require regulating such transmissions to avoid congestion and high latency. The paper investigates this offloading problem when transmissions regulation is through a token bucket, a mechanism commonly used for such purposes.  The goal is to devise a lightweight, online offloading policy that optimizes an application-specific metric (\eg classification accuracy) under the constraints of the token bucket.  The paper develops a policy based on a Deep Q-Network (DQN), and demonstrates both its efficacy and the feasibility of its deployment on embedded devices.  \add{Of note is the fact that the policy can handle complex input patterns, including correlation in image arrivals and classification accuracy.}
The evaluation is carried out by performing image classification over a local testbed using synthetic traces generated from the ImageNet image classification benchmark.  Implementation of this work is available at \url{https://github.com/qiujiaming315/edgeml-dqn}.

\end{abstract}


\section{Introduction}
\label{sec:intro}

Recent years have witnessed the emergence of \textit{Artificial Intelligence of Things (AIoT)}, a new paradigm of embedded systems that builds on two important advances. First, through progress in embedded hardware~\cite{almeida2019embench,ignatov2019ai,wang2020neural}, machine learning models can now run on embedded devices,
even if resource constraints limit them to relatively \textit{weak} models~\cite{sandler2018mobilenetv2,han2015deep,jacob2018quantization} that trade accuracy for resource efficiency. Second, edge servers accessible through shared local networks are increasingly common, providing access to additional compute resources~\cite{shi2016edge}.  Those edge servers are powerful enough to run \textit{strong(er)}, more complex models that are more accurate, therefore supplementing the weak local models running on embedded devices.
%
\begin{figure}
\centering
  \includegraphics[width=1\linewidth]{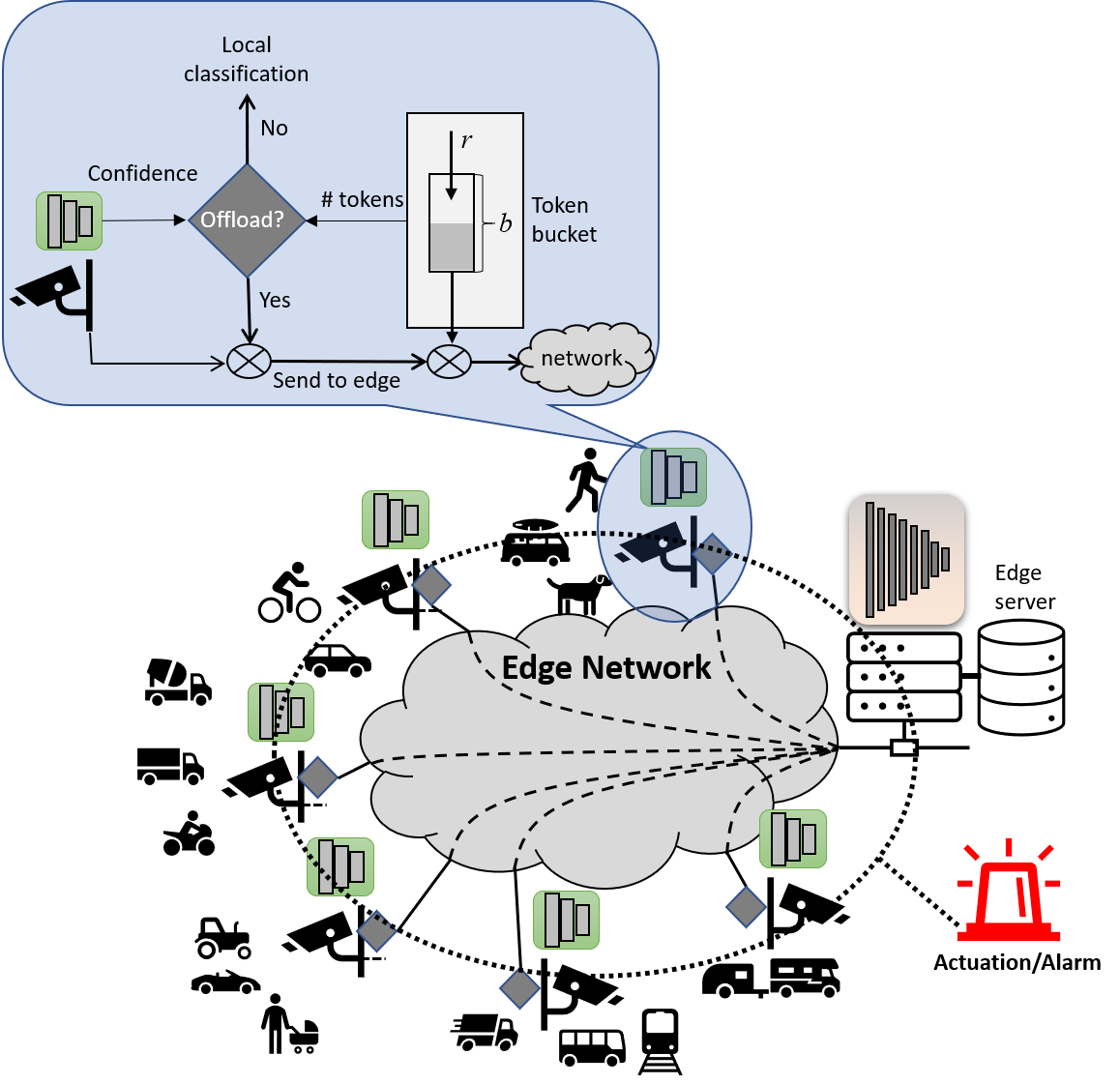}
  \caption{System overview and connectivity.}
\label{fig:oview}
\end{figure}%

\vspace{-4mm}
Of relevance in our setting is that independent of the edge compute resources, the large amount of input data (\eg images) acquired by embedded devices and the limited bandwidth of the shared network call for judicious decisions on what to \textit{offload} to edge servers and when. In particular, bandwidth constraints call for rate limiting transmissions from embedded devices.  In this work and following common practice, we employ a standard token bucket~\cite[Section 18.4.2]{medhi18} to regulate offloading traffic.  A token bucket (sometimes called a leaky bucket) provides a simple and flexible mechanism that specifies both a long-term transmission \textit{rate} and a maximum number of consecutive transmissions (\textit{bucket size}).  It has become the de facto standard for limiting user transmissions in both wired and wireless networks, with implementations available across commercial router/switch products, 
cloud providers offerings, 
and all major operating systems 
and programming languages. 
As a result, the findings of the paper should have applicability beyond the specific environment it considers.

\IEEEpubidadjcol

\fig{fig:oview} offers a representative example of the type of edge computing setting
we consider.  We use image classification as our target application, although the framework may be generalized to other types of classification or inference applications.

\IEEEpubidadjcol

Cameras distributed across an area 
share a network connecting them to an edge server.  They are responsible for capturing images and classifying them according to the category to which they belong.  As is common~\cite{krizhevsky2012imagenet}, this is done using a deep learning model. The limited computational resources available in the cameras impose the use of what we term a weak model in contrast to the strong model available on  the edge server that boasts greater compute resources.  The primary difference between the two models is the confidence metric of their outputs, with the strong model outperforming the weak one.  In many instances, the weak model returns a satisfactory (of sufficient confidence) answer, but it occasionally falls short.  In those cases, the embedded device has the option to send its input to the edge server for a higher confidence answer.  However, network bandwidth constraints call for regulating such offloading decisions through a token bucket mechanism, with each image transmission consuming a token. The challenge is to devise a policy that meets those constraints while maximizing classification accuracy (the metric of interest). 

Offloading decisions influence both immediate and future ``rewards'' (improvements in classification accuracy).  Offloading an image generates an immediate reward from the higher (expected) accuracy of the edge server classification.  However, the token this transmission consumes may be better spent on a future higher reward image.  This trade-off depends on \emph{both} future image arrivals and how the classifiers would perform on those images.  Neither aspect is likely to follow a simple pattern.  For example, image capture may be triggered by external events (\eg motion detectors), with the resulting arrival process exhibiting complex variations.  Similarly, the accuracy of the weak classifier may be influenced by weather and lighting conditions or the type of objects in the images.  This may in turn introduce correlation in the accuracy of consecutive images classifications.  

Examples of real-world image classification applications that may exhibit such complex input patterns include automatic check-out in retail stores, wildlife monitoring, or AI-powered robots that classify waste in recycling plants. In all those settings, external factors, \eg store layout, animals behavior, or how items are stacked in recycling bins, can produce complex input sequences to the classifier. 

This paper presents a general solution capable of handling arbitrary input sequences while making efficient offloading decisions on embedded devices. The solution is built on a Deep Q-Network (DQN) framework that can learn an efficient offloading policy given a training sequence of representative inputs, \ie based on a history of consecutive images, classification outputs, offloading rewards, and token bucket states.
More specifically, the paper makes the following contributions:
\begin{itemize}
\item A DQN-based policy that optimizes offloading decisions under variable image arrival patterns and correlation in the accuracy of consecutive images classifications, while accounting for token bucket constraints;
\item An implementation and benchmarking of the policy in an edge computing testbed demonstrating its efficiency on embedded devices;
\item A comprehensive evaluation using a wide range of image sequences from the ImageNet dataset, illustrating its benefits over competing alternatives.
\end{itemize}

\section{Background and Motivation}
\label{sec:background}


\begin{figure*}[!t]
\centering
    \subfloat[]{\includegraphics[width=0.2849\textwidth]{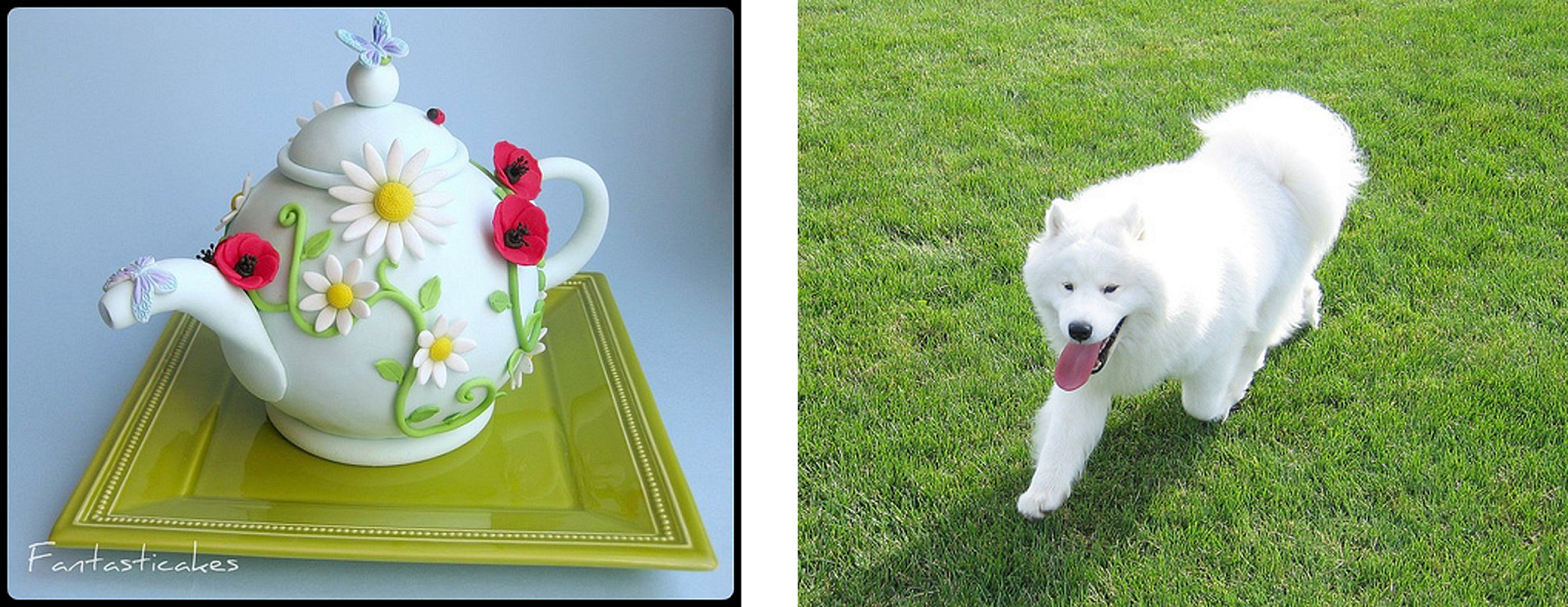}
    \label{fig:sub1a}}
    \hfil
    \subfloat[]{\includegraphics[width=0.2928\textwidth]{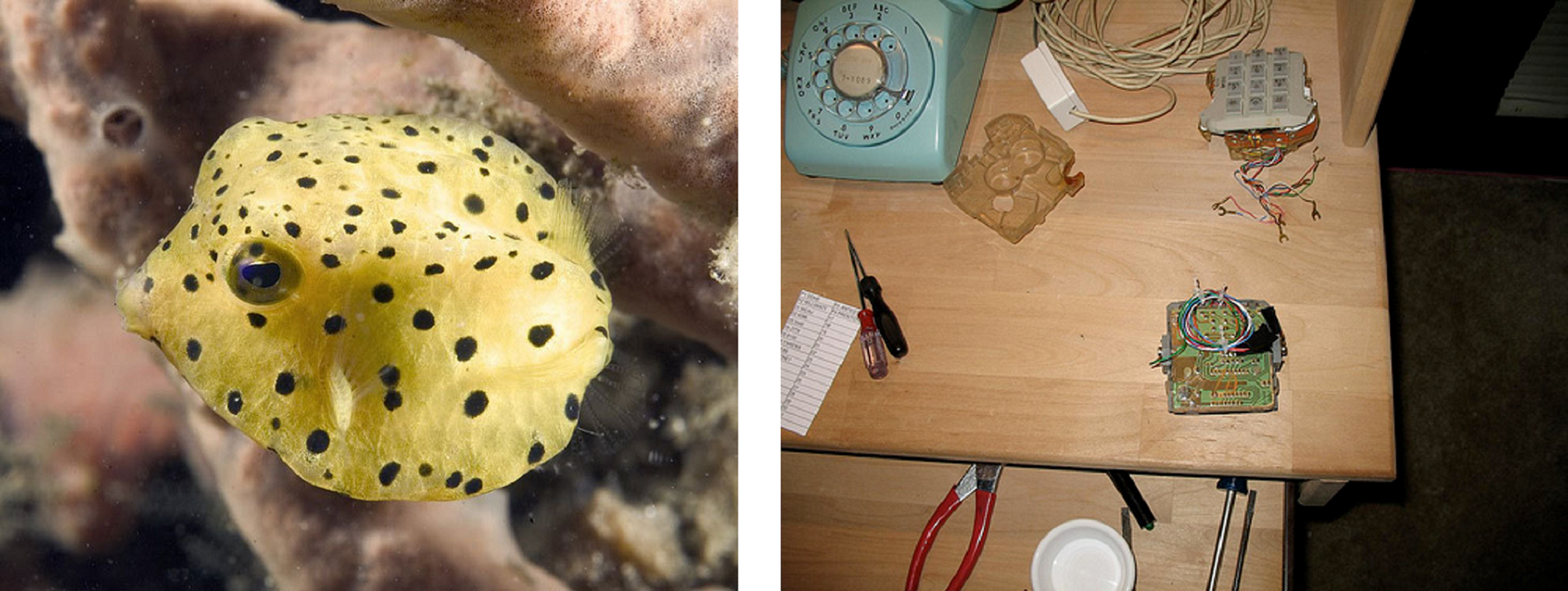}
    \label{fig:sub1b}}
    \hfil
    \subfloat[]{\includegraphics[width=0.3223\textwidth]{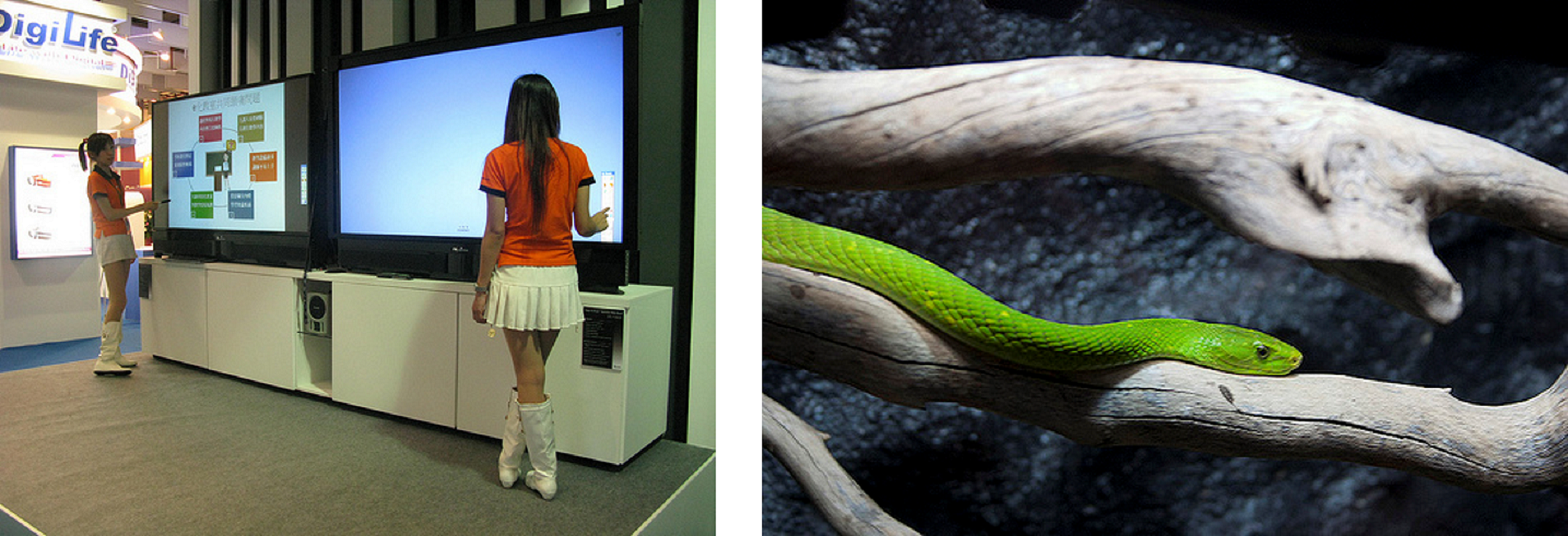}
    \label{fig:sub1c}}
    \caption{Image samples from the ILSVRC validation set for which classification is (a) accurate for both classifiers, (b) hard for both classifiers, and (c) accurate for the strong classifier but not the weak one.}
    \label{fig:example}
\end{figure*}

As mentioned in Section~\ref{sec:intro}, embedded devices can now run deep learning models. The co-location of data and processing offers significant benefits in leveraging distributed compute resources and timeliness of execution.  For example, as we report in Section~\ref{sec:cost}, local execution can return an image classification answer in about 20ms vs.~over 50ms if performed on an edge server after transmission over a local WiFi network.


This gain in timeliness, however, comes at a cost, as the weak(er) models running in embedded devices can under-perform the stronger models that edge servers can run.  Of interest though is the fact that differences in image classification accuracy are not systematic or even common.  Those differences vary depending on the classifiers (weak and strong) used, but broadly fall in three categories: (a) images that both classifiers accurately classify, (b) images that both classifiers struggle to classify accurately, and (c) images that the strong classifier can handle but not the weak classifier.

The relative fraction of images in each category can vary, but for typical combinations of classifiers many images are in (a), a small fraction of images are in (b), and the remainder are in (c).  For example, using the model of~\cite{cai2020once} with a computational footprint of~595MFlops as the strong classifier, and a $16$-layer VGG-style model as the weak classifier, we find that across the ILSVRC validation set 70.00\% of images are in (a), 4.47\% are in (b), and the remaining 25.53\% images are in (c) (\fig{fig:example} shows sample images from all three categories).

To improve overall classification accuracy, images in (c) should be offloaded, while offloading images in (a) or (b) is a waste of network bandwidth and edge resources.  Any solution must, therefore, first identify images in (c), and then ensure that as many of them can be transmitted under the constraints imposed by the rate control mechanism (token bucket). This is difficult because of the often unpredictable nature of the arrival pattern of images in (c). 
Developing a policy capable of handling this complexity
is one of the challenges the solution developed in this paper addresses.

\section{Related Work}

\add{\subsection{Edge Computing for Deep Learning Applications}}

Three general approaches have been explored to address bandwidth constraints in edge computing systems running deep neural network (DNN) models. We briefly review them.

\textbf{Input Adaptation:} In this approach the deep learning model is only deployed on the edge server, and the embedded devices offload \emph{all} inputs to the edge server for inference.  A variety of application-specific techniques have been exploited to reduce the size of the input data, including compression based on regions of interest (RoI) for object detection~\cite{ren2018distributed, liu2019edge}, adaptation of video frame size and rate~\cite{liu2018dare}, exploiting motion vector for object tracking~\cite{liu2019edge}, face cropping and contrast enhancement for emotion recognition~\cite{muhammad2021emotion}, and DNN-driven feedback regions for video streaming~\cite{du2020server}.  The key idea 
is to adapt the input as a function of the inference tasks towards preserving its accuracy. 
None of these solutions exploit the capabilities of modern embedded hardware to execute machine learning models locally. 


\textbf{Split Computing:} This approach takes advantage of the computing capability of embedded devices by splitting the inference task between the device and the server, with each side completing part of the computation.  The deep learning model is partitioned into head and tail models deployed on the device and the server, respectively. Early works~\cite{kang2017neurosurgeon,eshratifar2021jointdnn} partition the original DNN to minimize bandwidth utilization.
More recent techniques~\cite{eshratifar2019bottlenet,shao2020bottlenet++} modify the original DNN structure by injecting a bottleneck autoencoder that ensures a lightweight head model.  Other works~\cite{matsubara2019distilled, matsubara2020head} apply knowledge distillation techniques to train an autoencoder that serves as its head model and performs part of the inference task in addition to compressing the input. 
In all these solutions, the offloading rate is fixed once the splitting is selected.

\textbf{Model Cascade and Early Exiting:}  The cascade of models framework~\cite{wang2017idk,kouris2018cascade} relies on a \textit{cascade} of models of increasing complexity and accuracy to achieve fast and accurate inference with deep learning models.  A weak (and fast) model is used first, with stronger but computationally more expensive models invoked only if the weak model is not sufficiently confident of its output. In an edge computing setting, this naturally suggests deploying a pair of weak and strong models on embedded devices and servers, respectively~\cite{ran2018deepdecision,wang2018bandwidth}.
Distributed Deep Neural Networks (DDNN)~\cite{teerapittayanon2017distributed} have a similar focus but rely on \textit{early exiting}
to avoid redundant inferences.
Intermediate exits (\ie sub-branches) added to the DNN model allow inference queries to exit once confidence exceeds a threshold. 
As with the cascade framework, this readily maps to an edge computing setting by assigning early exit layers to the embedded device and the remaining layers to the edge server~\cite{laskaridis2020spinn,zeng2019boomerang}.  Of particular relevance is~\cite{laskaridis2020spinn} that seeks to select exit points based on network conditions.  However, none of those works focus on enforcing explicit \textit{rate limits} as imposed by token buckets.
\subsection{Computation Offloading Algorithms in Edge Computing}
\label{sec:computation_offloading}

Devising effective offloading policies is a fundamental problem in edge computing\footnote{\add{Lin \etal \cite{lin2020survey} provides a comprehensive review.
}}; one that has received significant attention.
In most works, the offloading problem is formulated as an optimization problem that aims to minimize a metric such as latency and/or energy consumption, with, as in this paper, deep Q-learning often the solution method of choice when dealing with dynamic and high-dimensionality inputs.

Focusing on a few representative examples, 
\cite{chen2019optimized} considers a mobile edge computing setup with sliced radio access network and wireless charging and relies on a double DQN approach
to maximize a 
utility function that incorporates latency and energy consumption.  Similarly,
\cite{min2019learning} investigates a scenario where energy harvesting IoT devices make 
offloading decisions across multiple edge servers and use DQN
to optimize offloading rate and edge server selection.  Finally,
\cite{huang2020deep} considers a wireless powered mobile edge computing system, and uses DQN
to make real-time offloading and wireless resource allocation decisions that adapt to 
channel conditions.

In spite of their reliance on DQN for offloading decisions in an edge computing setting, there are several important differences with this paper.  The first is that those papers aim to optimize general system or computational metrics rather than an application-specific metric (classification accuracy) that depends on \emph{both} local and edge performance.
In addition, although they also target an optimization under constraints, \eg energy constraints~\cite{chen2019optimized, min2019learning, huang2020deep},
those give rise to different state representations and, therefore, problem formulation than the token bucket constraint we consider.

The problem of optimizing offload decisions \add{to maximize inference accuracy} under token bucket constraint, \add{which we consider,} was \add{first} introduced in~\cite{chakrabarti2020real} based on the cascade of models framework. The work formulated the offloading decision problem as a Markov Decision Process (MDP) assuming that the inputs to the classifier are \textit{periodic} and \textit{independent and identically distributed (i.i.d.)}.  It generalized the fixed offloading threshold model of the cascade framework~\cite{wang2017idk,kouris2018cascade,teerapittayanon2017distributed,laskaridis2020spinn} to account for the token bucket constraints
by adopting an offloading policy that, for every token bucket state, learned a threshold based on the \add{local classifier} confidence score.
As alluded to in Section~\ref{sec:intro}, the periodic and i.i.d. assumptions may apply in some settings, but they are overly restrictive and unlikely to hold in many real-world applications.  Devising policies capable of handling \add{more} complex image sequences is the focus and main contribution of this paper. 
\section{Problem Formulation}

Recalling the system of 
\fig{fig:oview}, images captured by cameras are classified by the local (weak) classifier and an offloading decision is made based on that classifier's confidence and the token bucket state. 
This offloading policy can be formulated as an online constrained optimization problem that accounts for (i) the image arrival process, (ii) the output of the (weak) classifier, (iii) the token bucket state, and (iv) the metric to optimize (classification accuracy).  

In the rest of this section, we review our assumptions along each of those dimensions before formulating our optimization, with Section~\ref{sec:solution} introducing a possible solution suitable for the limited computational resources of embedded devices.

\subsection{Input Process}
\label{sec:input}

The first aspect affecting offloading decisions is \emph{how} inputs arrive at each device, both in terms of their frequency (rate) and temporal patterns.  Our goal is to accommodate as broad a set of scenarios as possible, and we describe next our model for the input arrival process at each device.  

For modeling sake, we assume a discrete time system with an underlying clock that determines when images can arrive.  Image arrivals follow a general inter-arrival time process with an arbitrary distribution $F(t)$.  This distribution can be chosen to allow both renewal and non-renewal inter-arrival times.  This includes i.i.d. arrival processes that may be appropriate when images come from a large set of independent sources, as well as non-renewal arrival processes, \eg MAP~\cite{map90}, that may be useful to capture environments where image arrivals follow alternating periods of high and low intensity.

In general, a goal of our solution will be to \emph{learn} the specific structure of the image arrival process, as captured by $F(t)$, and incorporate that knowledge into offloading decisions.

\subsection{Classifier Output}
\label{sec:classifier}

The weak and strong classifiers deployed in the devices and the edge server are denoted as $\mathcal{W}$ and $\mathcal{S}$ respectively.  For a given image $x$ they provide classification outputs $\mathcal{W}(x)$ and $\mathcal{S}(x)$ in the form of probability distributions over the (finite) set of possible classes $\mathbb{Y}$.  Given the ground truth class $y$ and the classifier output $z$ for an input image $x$, an application-specific loss (error) function $L(z, y)$ is defined that measures the mis-classification penalty (\eg $0$ if $y$ is among the $k$ most likely classes according to $z$ and $1$ otherwise, when the application is ``top-$k$'').  Loss is, therefore, dependent on whether or not an image is offloaded, and for image $x$ denoted as $L(\mathcal{W}(x), y)$ if it is not offloaded, and $L(\mathcal{S}(x), y)$ otherwise.

Note that at (offloading) decision time both $\mathcal{S}(x)$ and $y$ are unknown so that neither $L(\mathcal{W}(x), y)$ nor $L(\mathcal{S}(x), y)$ can be computed.  As a result and as discussed in Section~\ref{sec:policy}, the policy's goal is instead to maximize an \emph{expected} reward (decrease in loss) from offloading decisions.  This reward is affected not just by the input arrival process, but also by the classifier output process.  In particular, dependencies in the classifier outputs, \eg caused by changes in environmental conditions, can result in sequences of high or low confidence outputs that need to be accounted for by the policy's decisions.

\subsection{Token Bucket}
\label{sec:token}

As mentioned, it is necessary to regulate the offloading rate of devices to control the network load. This is accomplished through a two-parameters token bucket $(r,b)$ in each device, which controls both short and long-term offloading rates.

Specifically, tokens are replenished at a rate of $r\leq 1,$ (fractional) tokens per unit of time, and can be accumulated up to a maximum value of $b$.  Every offloading decision requires the availability of and consumes a full token.  Consequently, the token rate, $r$, upper-bounds the long-term rate at which images can be offloaded, while the bucket depth, $b$, limits the number of successive such decisions that can be made. 

Reusing the notation of~\cite{chakrabarti2020real}, the behavior of the token bucket system can be captured by tracking the evolution of the token count $n[t]$ in the bucket over time, as follows:
\beq
    n[t+1] = \min(b, n[t] - a[t] + r),
\eeq
where $a[t]$ the offloading action at $t$, which is $1$ if an image arrives and is offloaded (this needs $n[t]\geq 1$), and $0$ otherwise.

Again as in~\cite{chakrabarti2020real}, we assume that both $r$ and $b$ are rational so that $r=N/P$ and $b=M/P$ for some integers $N \leq P \leq M$.  We can then scale up the token count by a factor of $P$ and express it as $\bar{n}$:
\beq
    \bar{n}[t+1] = \min(M, \bar{n}[t] - P \times a[t] + N),
\eeq
which ensures that $\bar{n}[t]$ is an integer in the set $\{N,N+1,\cdots,M\}$, with images offloaded only when $\bar{n}[t] \geq P$.

\subsection{Offloading Reward and Decisions}
\label{sec:policy}

The offloading policy seeks to ``spend'' 
tokens on
images that maximize an application-specific metric (classification accuracy) while conforming to the token bucket constraints.

Suppose at time unit $t$ the image $x[t]$ with ground truth category $y[t]$ arrives, so that, as defined earlier, the loss of the classification predictions of the weak and strong classifiers are $L(\mathcal{W}(x[t]), y[t])$ and $L(\mathcal{S}(x[t]), y[t])$, respectively.  We define the offloading reward $R[t]$ as the reduction in loss through offloading the image to the edge:
\beq
\label{eq:reward}
R[t] = L(\mathcal{W}(x[t]), y[t]) - L(\mathcal{S}(x[t]), y[t]).
\eeq
Under the assumption of a general input process, a policy $\pi$ making an offloading decision $a[t]$ at time~$t$ may need to account for the entire input history up to time~$t$ as well as the scaled token count $\bar{n}[t]$, namely,
\beq\label{eq:policy}
    a[t] = \pi(X[t], \bar{n}[t]),
\eeq
where $X[t]$ is the input history from time~$0$ to time~$t$ that accounts for past image arrivals and classification outputs.

As alluded to in Section~\ref{sec:classifier}, we seek an offloading policy $\pi^*$ that maximizes the expected sum of rewards over an infinite horizon with a discount factor $\gamma \in [0, 1)$. In other words,
\beq\label{eq:opt_policy}
    \pi^* = \argmax_{\pi}\mathbb{E}\sum\limits_{t=0}^\infty\gamma^t a[t]R[t].
\eeq
Note that, when no image arrives at time $t$, we implicitly assume that $x[t]$ is null and that correspondingly so is the classification output. The offloading action $a[t]$ and reward $R[t]$ are then both $0$.  This ensures that the input history $X[t]$ incorporates information on past image inter-arrival times and the classification outputs following each image arrival, with the policy only making decisions at image arrival times.

\section{Solution}
\label{sec:solution}

We now describe the approach we rely on to derive
$\pi^{*}$.  The policy assumes a given pair of image classifiers $\mathcal{W}$, $\mathcal{S}$, access to representative training data, and seeks to specify actions that maximize an expected discounted reward as expressed in \Eqref{eq:opt_policy}.  There are several challenges in realizing $\pi^*$.

The first is that, 
to improve classification accuracy by taking advantage of the edge server's strong classifier, we need to identify images with a positive offloading reward (\ie images in (c) as described in Section \ref{sec:background}).
Based on \Eqref{eq:reward}, the reward associated with an input $x(t)$ depends on the outputs of \emph{both} the weak and strong classifiers, $\mathcal{W}(x[t])$ and $\mathcal{S}(x[t])$, \emph{and} knowledge of the true class $y(t)$ of the input.  Unfortunately, neither $\mathcal{S}(x[t])$ nor $y(t)$ are available at the time an offloading decision needs to be made.  We address this challenge through an approach similar to that of~\cite{chakrabarti2020real} that relies on an offloading metric $m\left(x\right)$, 
which \emph{learns} an estimate of the offloading reward $R[t]$.
We briefly review this approach in Section~\ref{sec:metric}.

The second more significant challenge is that, as reflected in \Eqref{eq:policy}, policy decisions may need the entire history of inputs (and associated metrics) to accurately capture dependencies in arrival patterns and classification outputs.  The size of the resulting state space can translate into significant complexity, which we address through a deep reinforcement learning approach based on \emph{Q-values} as in~\cite{mnih2013playing}. We expand on this approach in Section~\ref{sec:dqn}.

In summary, the processing pipeline for each image in an embedded device has following steps:  (1) The weak classifier classifies the image and produces an output $\mathcal{W}(x)$; (2) Using $\mathcal{W}(x)$ the offloading metric $m(x)$ is computed as an estimate of the reward $R$; (3) $\mathcal{Q}$-values are then computed based on the current state (which includes a history of offloading metrics and input inter-arrival times, and the token bucket state) and an offloading decision is made. Of note is that $\mathcal{Q}$-values rely only on current and local information, which allows for timely offloading decisions independent of the edge server.

\subsection{Offloading Metric}
\label{sec:metric}

As mentioned, each time an image $x$ arrives, the only information available after its local processing is the output of the weak classifier $\mathcal{W}(x)$.  The offloading metric $m(x)$ represents then an estimate for the corresponding offloading reward $R$.  
We compute $m(x)$ following the approach outlined in~\cite[Section 4.1]{chakrabarti2020real}, which uses a training set of $K$ representative image samples to generate a mapping from the entropy $h\left(\mathcal{W}(x)\right)$ of the weak classifier output to the expected reward.

The entropy $h(z)$ of a classification output $z$ is given by: 
\bequn
    h(z) = -\sum_{y \in \mathbb{Y}}z_{y}\log z_{y},
\eequn
which captures the classifier's confidence in its result (recall that the classifier's output is in the form of a probability distribution over the set of possible classes).  This entropy is then mapped to an expected offloading reward using a standard radial basis function kernel:
\beq
\label{eq:mapping}
    f(\bar{h}) = \frac{\sum_{k=1}^{K}\sigma(\bar{h}, h_k)\times R_k}{\sum_{k=1}^{K}\sigma(\bar{h}, h_k)},
\eeq
where $\bar{h}=h(z)$ for classification output $z$, $\sigma(\bar{h}, h_k) = \exp\left(-\lambda(\bar{h} - h_k)^2\right)$, and $R_k$ is the reward from the $k^{th}$ sample in the training set with $h_k$ its entropy.

By setting $m(x) = f(h(\mathcal{W}(x)))$, we choose an expected reward that is essentially a weighted average over the entire training set of $K$ images of reward values for training set inputs with similar entropy values, where images with entropy values closer to that of image $x$ are assigned higher weights.

\subsection{A Deep Q-Learning Policy}
\label{sec:dqn}

With the metric $m(x)$ of image $x$ in hand, the policy's goal is to decide whether to offload it given also the system state as captured in $X(t)$ and $\bar{n}(t)$, the past history of image arrivals, classification outputs, and the token bucket state.  The potential sheer size of the underlying state space makes a direct approach impractical. This leads us to exploring the use of deep Q-learning proposed in~\cite{mnih2013playing}.  In the remainder of this section, we first provide a brief overview of deep Q-learning before discussing its mapping to our problem and articulating its use in \emph{learning} from our training data set an offloading policy that seeks to maximize the expected offloading reward.

\subsubsection{Background}

\emph{Q-learning} is a standard Reinforcement Learning approach for devising policies that maximize a discounted expected reward summed over an infinite horizon as expressed in \Eqref{eq:opt_policy}.  It relies on estimating a \emph{Q-value}, $\mathcal{Q}(s, a)$ as a measure of this reward, assuming that the current state is $s$ and the policy takes action $a$.  As mentioned above, in our setting, $s$ consists of the arrival and classification history $X$ and the token count $\bar{n}$, while $a$ is the offloading decision.

Estimating Q-values relies on a \emph{Q-value function}, which in \emph{deep Q-learning} is in the form of a deep neural network, or Deep Q-Network (DQN).  Denoting this network as $\mathcal{Q}$, it learns Q-values during a training phase through a standard \emph{Q-value update}. Specifically, denoting the current DQN as $\mathcal{Q}^-$ let
\beq\label{eq:q_update}
    \mathcal{Q}^{+}(s, a) = R(s,a,s') + \gamma\max_{a'}\mathcal{Q}^{-}(s', a'),
\eeq
where $s'$ is the state following action $a$ at state $s$, $R(s,a,s')$ is the reward from this transition (available during the training phase) with $\gamma$ the discount factor of \Eqref{eq:opt_policy}, and both $a$ and $a'$ are selected from the set of feasible actions in the corresponding states $s$ and $s'$. 

The value $\mathcal{Q}^{+}(s, a)$ is used as the ``ground-truth'', with the difference between $\mathcal{Q}^{+}(s, a)$ and $\mathcal{Q}^{-}(s, a)$ representing a loss function to minimize, which can be realized by updating the weights of the DQN through standard gradient descent.  The approach ultimately computes Q-values for all combinations of inputs (state~$s$) and possible actions~$a$, and the resulting policy greedily takes the action with maximum Q-value in each state:
\beq\label{eq:q_policy}
    \pi(s) = \argmax_{a}\mathcal{Q}(s, a).
\eeq
The challenges in learning the policy of \Eqref{eq:q_policy} are the size of the state space and the possibility of correlation and non-stationary input distributions, which can all affect convergence. Deep Q-learning introduced two additional techniques to address those challenges:

\noindent\textbf{Experience replay:} The Q-value updates of \Eqref{eq:q_update} rely on a $(s, a, R, s')$ tuple, where we recall that the state $s$ may include the entire past history of the system, \eg the tuple $(X,\bar{n})$ of \Eqref{eq:policy} in our case.  Deep Q-learning generates (through simulation\footnote{As we shall see shortly, our setting mostly avoids simulations.}) a set of $(s, a, R, s')$ tuples, stores them in a so-called \emph{replay buffer}, which it then randomly samples to perform Q-value updates.  This shuffles the order of the collected tuples so that the learned Q-values are less likely to diverge because of bias from groups of consecutive tuples.

\noindent\textbf{Target network:} A Q-value update changes the weights of the DQN and consequently its Q-value estimates in subsequent updates.  Deep Q-learning makes a separate copy of the DQN, known as the \emph{target network}, $\mathcal{Q}_{target}$, which it then uses across multiple successive updates. Specifically, the Q-value update of \Eqref{eq:q_update} is modified to use:
\beq\label{eq:DQN_update}
    \mathcal{Q}^+(s, a) = R(s, a, s') + \gamma\max_{a'}\mathcal{Q}_{target}(s', a').
\eeq
Weights of the current DQN are still modified using gradient descent after each update, but subsequent values continue to be computed using $\mathcal{Q}_{target}$.  The two networks are eventually synchronized, \ie $\mathcal{Q}_{target}$ is updated to the current DQN, but limiting the frequency of such updates has been shown to improve learning stability. 

\subsubsection{DQN Setup}
\label{sec:dqn_setup}
This section introduces the architecture and setup of the DQN used to estimate Q-values for making efficient offloading decisions based on the structure of the input process, dependencies in the classification output, and the token bucket state.  Aspects of relevance to our DQN include its inputs and outputs, as well as its internal architecture.

Our system state consists of the input $X$ (image arrivals and classification history) and the (scaled) token count $\bar{n}$, \ie $s=(X,\bar{n})$.  
For computational efficiency,
rather than using raw images, we instead rely on the offloading metrics $m(x)$ to estimate Q-values\footnote{
Using raw images would add a component of complexity comparable to the weak classifier itself, which is undesirable. An alternative is to use intermediate features extracted from the weak classifier. This is still challenging, especially when considering a history of such metrics, as the dimensionality of these features remains much higher than the offloading metric (a scalar), and would likely require a more complex model architecture.}.
The input history $X$ therefore reduces to $(\mathbf{I}, \mathbf{m})$, \ie the history of image inter-arrival times and offloading metrics.  As mentioned earlier, the state space is independent of the strong classifier, so that offloading decisions can be made immediately based only on local information.

With this state definition, Q-values are produced for each combination of $(X,\bar{n},a)$,
where $a$ is a (feasible) offloading decision.  This suggests $(X,\bar{n},a)$ as our input to the DQN.  Such a selection is, however, relatively inefficient; both from a runtime and a training standpoint.  From a runtime perspective, it calls for multiple passes through the DQN, one for each possible action. More importantly, a different choice of input can significantly improve \emph{training} efficiency.

In particular, token states are a deterministic function of offloading actions and our inputs (and metrics) are statistically independent of actions. This allows the parallel computation of Q-values across possible actions, and computing (and updating during the training phase) Q-values for \emph{all} token bucket states $\bar{n}$ at the same time without resampling training data based on policy, \ie avoid doing proper reinforcement learning.  This can significantly improve training efficiency.  As a result, we select $X$ as our system \emph{input}, with our DQN producing a set of $2M-P-N+2$ \emph{outputs} (Q-values), one for each combination of token bucket states
$\bar{n}$ and offloading actions $a\in\{0,1\}$.

Many recent works in deep reinforcement learning
involve relatively complex deep convolutional neural networks (CNN) to handle high-dimensional inputs such as raw images, 
or rely on more sophisticated algorithms than DQN, \eg Proximal Policy Optimization (PPO)~\cite{schulman2017proximal} or Rainbow~\cite{hessel2018rainbow}.
Initial experiments with CNNs did not yield meaningful improvements over a lightweight multi-layer perceptron (MLP), possibly
from our state space relative low dimensionality. As a result, given our focus on a light computational footprint, we opted for a simple MLP architecture with $5$ layers and $64$ units in each layer\footnote{
The performance impact of different choices is discussed in Section~\ref{sec:dqn_param}.},
and the relative simplicity of the DQN algorithm.  Exploring the feasibility and benefits of more sophisticated RL algorithms and more complex architectures such as recurrent neural networks (RNN) is a topic we leave to future work.

\subsubsection{DQN Learning Procedure}
As our inputs $X$ are independent of actions and the token state is a deterministic function of action, we can limit ourselves to generating a sequence of image arrivals and corresponding the offloading metrics and rewards as our training set, which we store in our \emph{replay buffer}.  

During training, the replay buffer is randomly sampled, each time extracting a finite history window (segment) of length $T$, which is assumed sufficient to allow learning the joint distribution of inter-arrival times and classification outputs.  Segments sampled from the beginning of the image sequence are zero-padded to ensure a window size of $T$ for all segments.  For each segment, we create an input tuple $X= (\mathbf{I}, \mathbf{m})$ that consists of the \emph{first} $T-1$ image inter-arrival times and the corresponding offloading metrics.  Conversely, the tuple $X'$ includes the same information but for the \emph{last} $T-1$ entries in the segment, and represents our next ``input state''.  We can then adapt the Q-value update expression of \Eqref{eq:DQN_update} as follows:
\beq\label{eq:DQN_update1}
    \mathcal{Q}^+(X,\bar{n}; a) = a \cdot R + \gamma\max_{a'\in\{0, 1\}}\mathcal{Q}_{target}(X', \bar{n}'; a'),
\eeq
where $\bar{n}$ is the token state when the current image (last entry in $X$) arrives, $R$ is the reward from offloading it, $a$ is the offloading decision for that image ($a$ is $0$ when $\bar{n} < P$), and $\bar{n}'$ is the updated token state following action~$a$.  Note that since no additional images can be offloaded until the next one arrives, $\bar{n}'$ can be readily computed from $\bar{n}, a$, and the last inter-arrival time $I_T$ in $X'$, namely, 
\bequn
    \bar{n}' = \min(M, \bar{n} - P \times a + N \times I_T),
\eequn
This also means that for any pair $(X,X')$ from a given  segment in our replay buffer, we can simultaneously update \emph{all} $\mathcal{Q}$-values associated with different token states. This significantly speeds-up convergence of our learning process.

\section{Evaluation}
\label{sec:evaluation}

Our goal is to demonstrate that the DQN-based policy (i) estimates Q-values efficiently with negligible overhead in embedded devices, and (ii) can learn complex input structures to realize offloading decisions that outperform state-of-the-art solutions.
To that end, we implemented a testbed emulating a real-world edge computing setting, and, in addition to simulations, ran extensive experiments to evaluate the policy's  runtime efficiency on embedded devices and its performance for different configurations.
Section~\ref{sec:setup} reviews our experimental setup. Section~\ref{sec:cost} presents our implementation and empirical evaluation of runtime efficiency in embedded systems. Finally, Section~\ref{sec:results} evaluates our policy's efficacy in making offloading decisions for different input structures.

\subsection{Experimental Setup}
\label{sec:setup}

\subsubsection{Classification Task}  We rely on the standard task of image classification with $1000$ categories from the ImageNet Large Scale Visual Recognition Challenge (ILSVRC) to evaluate the classification performance of our offloading policy.

Our classification metric is the \emph{top-5 loss} (or error). It assigns a penalty of $0$ if the image is in the five most likely classes returned by the classifier and 1 otherwise.  The strong classifier in our edge server is that of~\cite{cai2020once} with a computational footprint of~595MFlops.  Our weak classifier is a ``home-grown'' $16$ layers model acting on low-resolution $64\times 64$ images with $13$ convolutional layers (8 with $1\times 1$ kernels and $5$ with $3\times 3$ kernels) and $3$ fully connected layers.

\begin{figure}[t]
	\centering
	\includegraphics[width=0.75\linewidth]{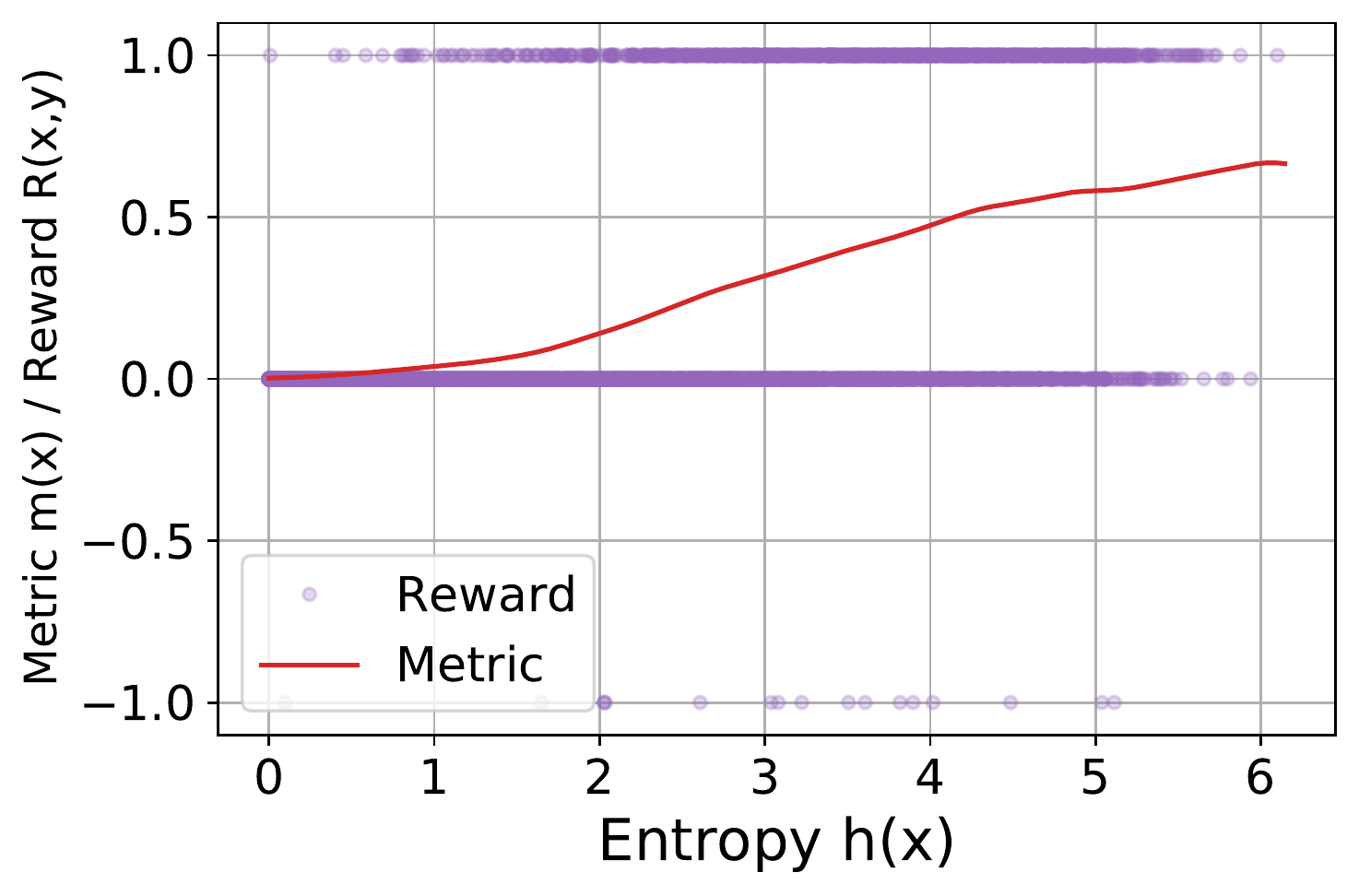}
    \caption{Mapping (red curve) from entropy of weak classifier output to offloading metric, with \emph{actual} rewards for training set images (purple dots).}
	\label{fig:offloading_metric}
\end{figure}

Given our classifiers and the top-5 loss metric, the function $f(h)$ of \Eqref{eq:mapping} that maps the entropy\footnote{Prior to computing the entropy, we calibrate the predictions of the weak classifier using temperature-scaling as outlined in \cite{guo2017calibration}. } of the weak classifier output to the offloading rewards across the training set is reported in \fig{fig:offloading_metric}.  We note that the relatively low prediction accuracy of our weak qualifier results in a monotonic mapping from entropy to metric, \ie in most instances where the weak classifier is very uncertain about its decision, the strong classifier can provide a more confident (and accurate) output.

\subsubsection{Image Sequence Generation}
The other main aspect of our experimental setup is our ``image generators.'' They determine both the image arrival process and how those images are sampled from the ImageNet dataset.  The former affects temporal patterns in image arrivals at the weak classifier, while the latter determines potential similarities among successive classification outputs.  To test our solution's ability to infer such patterns, distinct sequence generators separately control image arrivals and similarities in classification outputs.

\subsubsection*{Image Arrival Process}\label{sec:arrival}  We rely on a simple two-state Markov-Modulated mechanism to create variable image arrival patterns.  Each state is associated with a different but fixed image inter-arrival time, $I_1$ and $I_2$, with each state having a given probability $tprob_i, i=1,2$, of transitioning to the other state. Given our discrete-time setting, up to one image arrives in each time slot, and the two states emulate alternating periods of high and low image arrival rates.
Of interest is the extent to which DQN recognizes when it enters a state with a lower/higher image arrival rate and adjusts its offloading decisions based not only on the token bucket state but also its estimate on when the next images might arrive.

\subsubsection*{Image Selection Process}  In the simplest instance, images are selected randomly from the ImageNet dataset.  This results in classification outputs with metrics randomly distributed across the ImageNet distribution.  As mentioned in Section~\ref{sec:classifier}, this may not be reflective of many practical situations.  
To create patterns of correlated confidence outputs, we rank-order the ImageNet dataset by images' offloading metric, and sample it using a simple two-parameter model based on a sampling spread $sp$ and a location reset probability $rprob$.  The reset probability $rprob$ determines the odds of jumping to a new random location in the rank-ordered ImageNet dataset, while the spread $sp$ identifies a range of images, and therefore metrics, from which to randomly select once at a location.  Correlation in the metrics of successive classification outputs can then be varied by adjusting $sp$ and $rprob$.

\subsubsection{DQN Configuration}  We use the official ILSVRC validation set with $50000$ images (1000 categories with 50 images each).  We evenly split the validation set into three subsets; two are used as training sets and the third as test set.  Given a token bucket configuration and sequence generator settings, we generate a training sequence of $10^8$ images from the training sets along with corresponding inter-arrival times and metrics.  This sequence is stored in the replay buffer from which we randomly sample (with replacement) input history segments with a fixed length history window of $T=97$ to train DQN.  The effect of the history window length $T$ on DQN's performance is investigated in Section~\ref{sec:dqn_param}.
Throughout the training procedure, we synchronize the target network with DQN every $2^{14}$ segments, and perform 4000 synchronizations, for a total of $4000 \times 2^{14} \approx 6.55\times 10^7$ segments for Q-value updates.  The DQN policy is then evaluated with test sequences of $10^7$ images from the test set sampled using the same sequence generator settings.

\subsubsection{Evaluation Scenarios} 
In evaluating DQN, we vary image arrival patterns, classification output correlation, and token bucket parameters, and compare DQN to several benchmarks.

The first is a \emph{lower bound} that corresponds to a setting where the weak classifier is limited to only offloading a fixed fraction of images based on its token rate $r$ (\ie images with offloading metrics above the $(1-r)^{th}$ percentile), but it is not constrained by the bucket size (equivalent to an infinite bucket size).  This lower bound is often not feasible, but barring knowing an optimal policy, it offers a useful reference.  

We also compare DQN to two practical policies.  The first is the \emph{MDP} policy introduced in~\cite{chakrabarti2020real}. It is oblivious to any structure in either the image arrival process or the classifier output (it assumes that they are i.i.d.), but is cognizant of the token bucket state and attempts to adapt its decisions based on the number of available tokens and its estimate of the long-term image arrival rate.  The second, denoted as \emph{Baseline}, is a \emph{fixed} threshold policy commonly adopted by many works in the model cascade 
framework~\cite{wang2017idk,kouris2018cascade,teerapittayanon2017distributed,laskaridis2020spinn}. \emph{Baseline} uses the same threshold as \emph{lower bound}, \ie attempting to offload images with offloading metrics above the $(1-r)^{th}$ percentile, but in contrast to \emph{lower bound}, it needs to conform to the token bucket constraint at run time. Further, unlike DQN, it is oblivious to the token bucket state and any structure in either the arrival process or the classification output.

\subsection{Runtime Efficiency}
\label{sec:cost}

\begin{figure*}[t]
	\centering
	\includegraphics[width=0.65\linewidth]{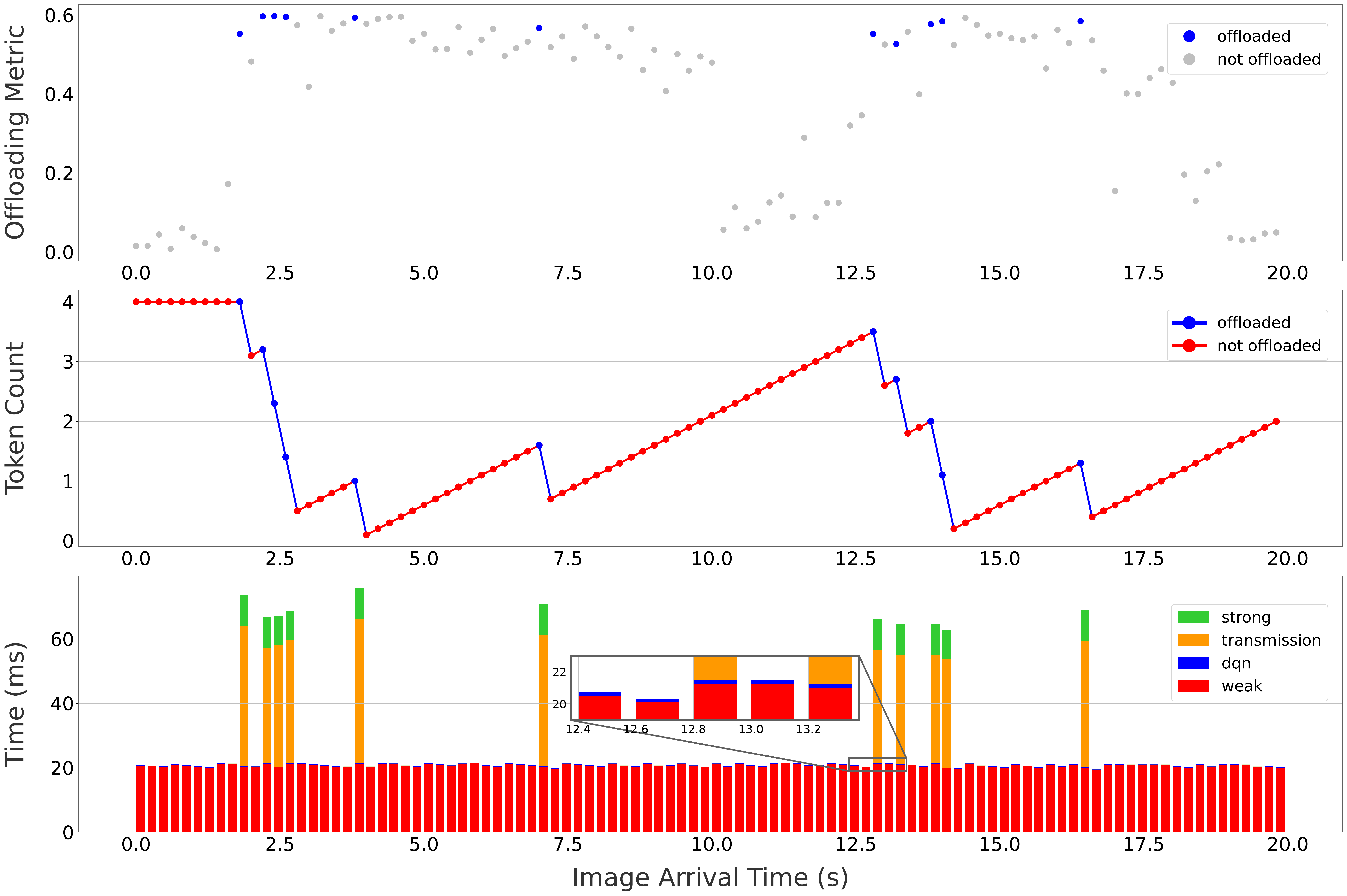}
	\caption{Traces of offloading metrics, token bucket states, and time spent in the image classification pipeline in a representative experiment.}
	\label{fig:testbed}
\end{figure*}

\begin{table*}[t]
\renewcommand{\arraystretch}{1.3}
\begin{center}
\caption{Time spent across components in the image classification pipeline} \label{tab:cost} 

\begin{tabular}{llcccc}
\hline
 \multicolumn{2}{l}{\textbf{Time} }& \textbf{Weak Classifier} & \textbf{DQN} & \textbf{Transmission} & \textbf{Strong Classifier} \\ 
 \hline
\multicolumn{2}{l}{\textbf{Absolute: mean(std) (ms)}} & $20.62 (0.57)$ & $0.25 (0.07) $ & $40.97 (9.30)$ & $11.64 (5.79) $ \\
 \hline
\multirow{2}{28pt}{\textbf{Relative:}}& \textbf{(not offloaded)} & $98.78\%$ & $1.22\%$ & $-$ & $-$\\ 
 &\textbf{(offloaded)} & $27.96\%$ & $0.33\%$ & $55.84\%$ & $15.87\%$ \\ 
 \hline
\end{tabular}
\end{center}
\end{table*}

To evaluate the feasibility of our DQN-based policy,
we implemented it on a testbed consisting of an embedded device and an edge server connected over WiFi, and quantified its overhead by comparing its runtime execution time on the embedded device to the time spent in other components in an end-to-end classification task.  Next, we briefly describe our testbed and measurement methodology. 

\subsubsection{Testbed Configuration}
\label{sec:testbed}

Our testbed comprises a Raspberry Pi 
4 Model B 8GB that costs $\sim$\$75
as the embedded device and a server equipped with an Intel(R) Core(TM) i7-10700K CPU @ 3.80GHz and Nvidia GeForce RTX 3090 GPU as the edge server.  The pair of weak and strong classifiers of Section~\ref{sec:setup} are deployed on the embedded device and the edge server, respectively.  To further accelerate the inference speed of the weak classifier, we convert the weak classifier to an 8-bit quantized TensorFlow Lite model and accelerate the inference with a Coral USB accelerator.  The DQN is also converted to a float16 TensorFlow Lite model.  The Raspberry Pi and the edge server communicate over a WiFi network using the 802.11/n mode from the 2.4GHz frequency band.

We resize the ILSVRC validation images to $236\times236$ in the pre-processing stage to unify the input images size to $1.34\times10^6$ bits, and set the image arrival rate to 5 images/sec.  To introduce correlation in consecutive classifications, we use $sp=0.1$ and $rprob=0.1$ for the classifier output process.

The token bucket is configured with a rate $r=0.1$ (\ie a long-term offloading rate of one out of 10 images or $0.67$~Mbps) and a bucket size $b=4$ (\ie allowing the offloading of up to 4 consecutive images).
We note that while the rate of $0.67$~Mbps is well below the bandwidth of the WiFi network, that bandwidth would in practice be shared among many embedded devices, so that rate controlling their individual transmissions, as we do, would be required.

\subsubsection{Computation Cost}

To quantify the overhead that DQN imposes, we measure where time is spent across the different components of the classification pipeline. 
The embedded device first classifies every image using its weak classifier, and then executes the DQN model to estimate the Q-values before making an offloading decision that accounts for the current token bucket state.  Offloaded images are transmitted to the edge server over the network and finally classified by the strong classifier.  Hence, a full classification task includes four main stages, (i) weak classifier inference, (ii) DQN inference, (iii) network transmission, and (iv) strong classifier inference, which all contribute to how long it takes to complete.

The bottom section of \fig{fig:testbed} plots those respective time contributions for a representative experiment involving a sequence of 100 images, with the two other sections of the figure reporting the metrics computed by DQN for each image (top) and the corresponding token counts (middle) and offloading decisions.
As we detail further in the rest of the section, the results illustrate how DQN takes both the offloading metric of each image and the token bucket state into account when making offloading decisions.  

As shown in Table~\ref{tab:cost}, DQN only takes 0.25~ms on average. This is just over 1\% of the time spent in the weak classifier, and for offloaded images, it is less than a third of a percent of the total classification pipeline time.  This demonstrates that the benefits DQN affords impose a minimal overhead.  Quantifying those benefits is the focus of the next section.


\subsection{Policy Performance}
\label{sec:results}

In this section, we evaluate DQN's performance across a range of scenarios, which illustrate its ability to learn complex input structures and highlight how this affects its offloading decisions. To that end we proceed in three stages.  In the first two, we introduce complexity in only \emph{one} dimension of the input structure, \ie correlation is present in either classification outputs or image arrivals.  This facilitates developing insight into how such structure affects DQN's decisions.  In the third stage, we create a scenario with complexity in \emph{both} classification outputs and image arrivals, and use it to demonstrate DQN's ability to learn policies when complexity spans multiple dimensions.  Finally, as a sanity check, we evaluate how different choices of 
model parameters, including history window length $T$, number of hidden layers, and number of units in each layer,
affect the performance of DQN.

%

\subsubsection{Deterministic Image Arrivals and Correlated Classification Outputs}
\label{sec:det+corr}
\begin{figure}[t]
	\centering
	\includegraphics[width=\linewidth]{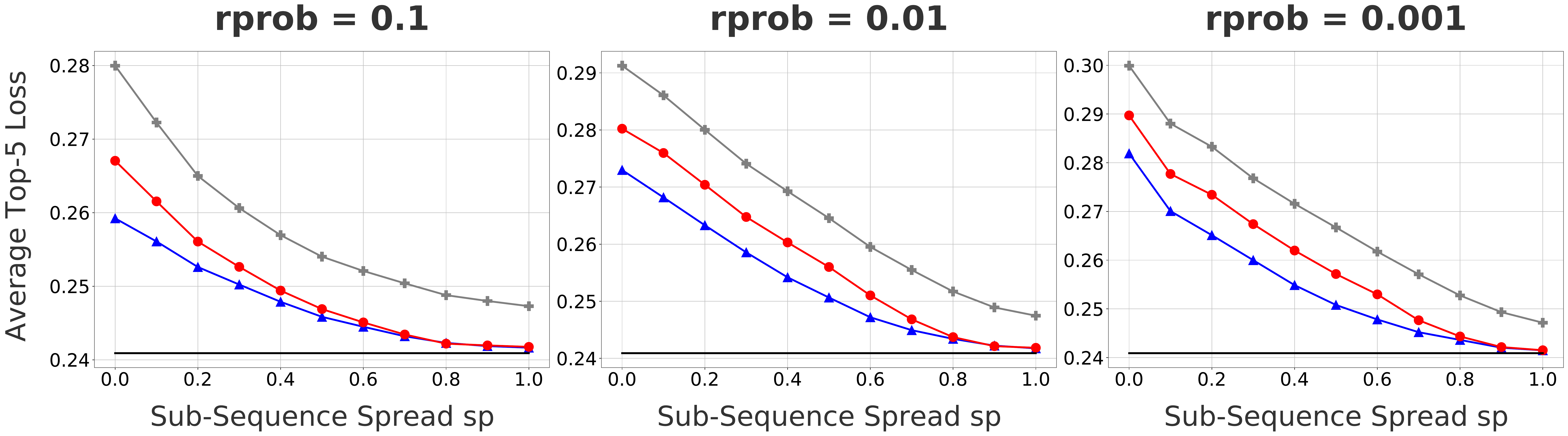}
	\includegraphics[width=\linewidth]{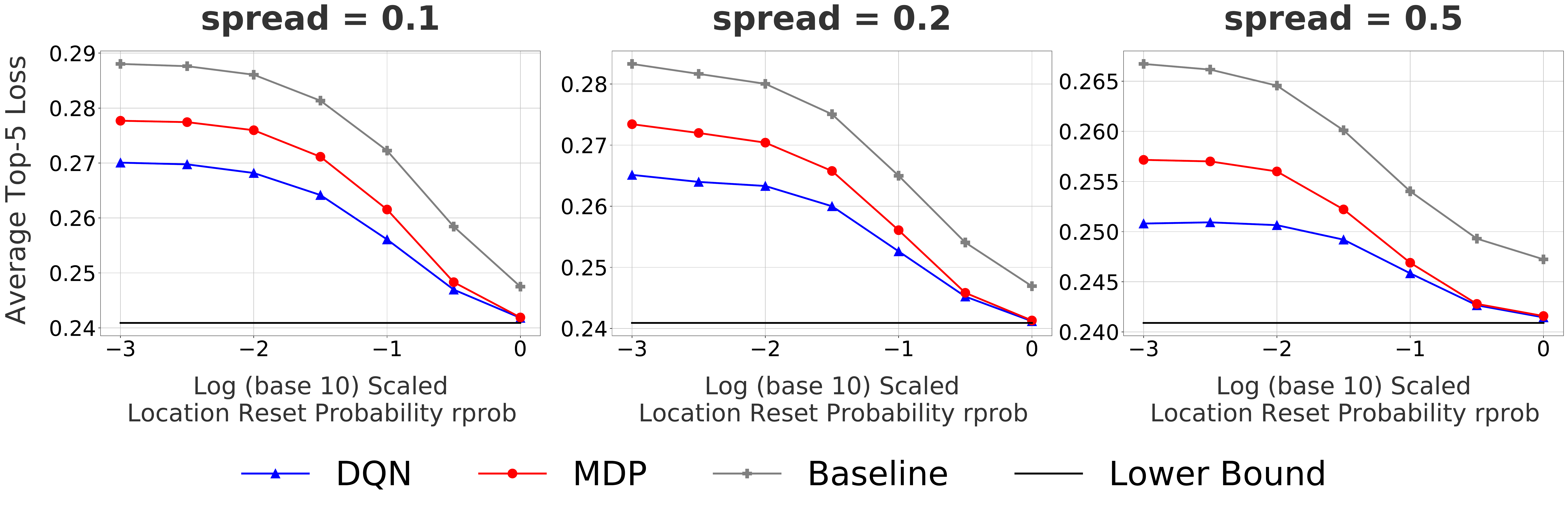}
	\caption{Offloading policies performance as a function of classifier output correlation.  Correlation decreases as spread $sp$ (Top) or location resetting probability $rprob$ (Bottom) increase. Token bucket: $r=0.1, b=4$.
	}
	\label{fig:output_correlation}
\end{figure}

To explore DQN's ability to learn about the presence of correlation in \emph{classification outputs,} we first fix the token bucket parameters to $r=0.1$ and $b=4$, and vary the two hyper-parameters of our sequence generator to realize different levels of classification output correlation: The sampling spread $sp$ is varied from $0$ (single image) to $1$ (full dataset and, therefore, no correlation), while the reset probability $rprob$ is varied from $10^{-3}$ to~$1$ (no correlation). \fig{fig:output_correlation} reports the top-5 loss for DQN and our three benchmarks. 

As expected, when either $sp$ or $rprob$ are large so that classification output correlation is minimal, both DQN and MDP perform similarly and approach the performance of the lower bound.  However, when classification output correlation is present, DQN consistently outperforms MDP (and the Baseline).  As correlation increases, performance degrades when compared to the lower bound, but this is not surprising given the token bucket constraints.  Correlation in the classification output means that sequences of either high or low metrics are more likely, which are harder to handle under token bucket constraints.  A sequence of high metric images may rapidly deplete a finite token bucket, so that it may not be possible to offload all of them, irrespective of how forward looking the policy is.  Conversely, a sequence of low metric images may result in wasted tokens (the bucket fills up) even if, as we shall see, the DQN policy is able to mitigate this by recognizing that it has entered such a period and adapting its behavior.

\begin{figure}[t]
	\centering
    \includegraphics[width=\linewidth]{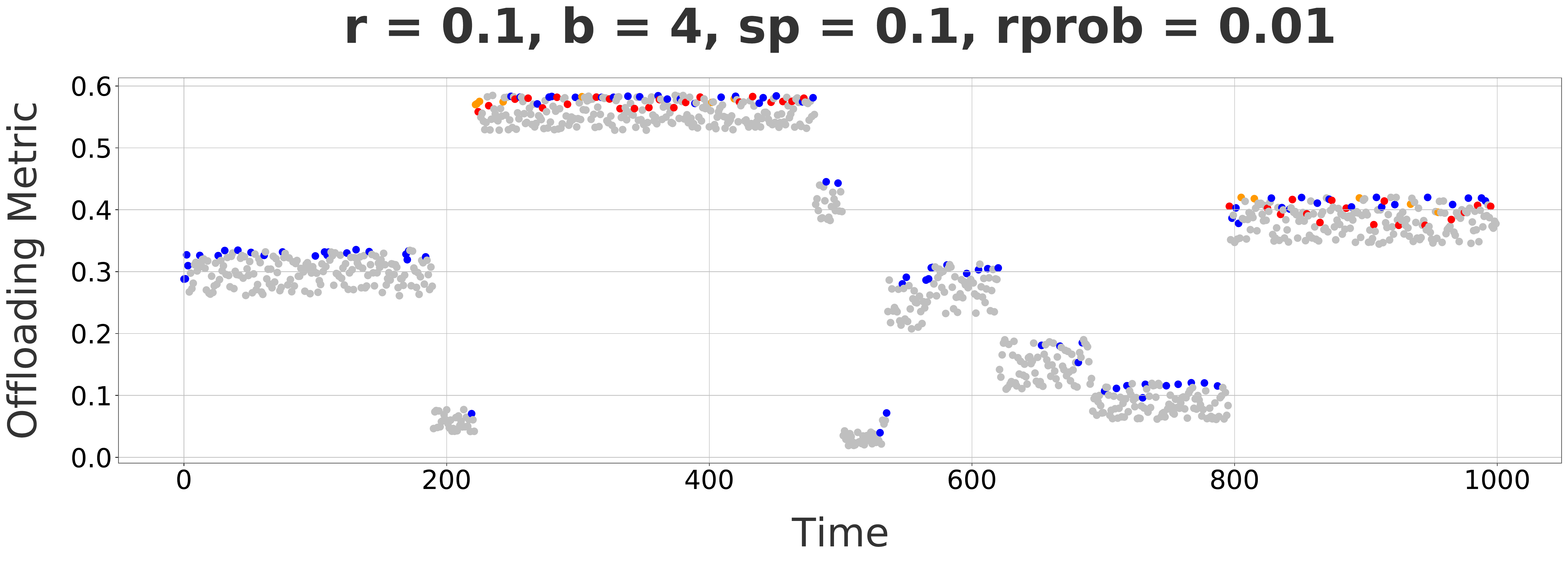}
	\includegraphics[width=\linewidth]{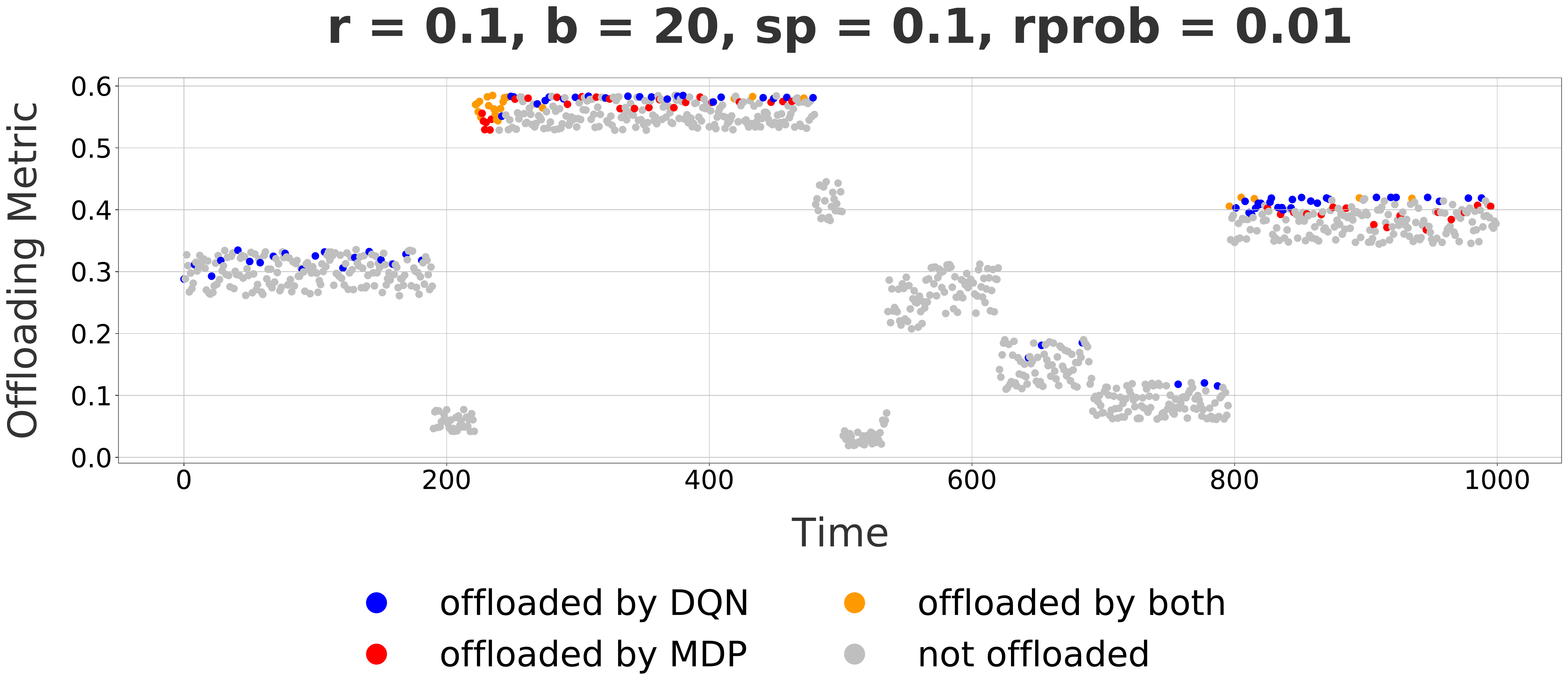}
	\caption{Offloading decisions of DQN and MDP for token bucket depths of $b=4$ (Top) and $b=20$ (Bottom) under correlated classification outputs ($sp=0.1$ and $rprob=0.01$).}
	\label{fig:output_depth_trace}
\end{figure}

This is illustrated in the top portion of \fig{fig:output_depth_trace} that reports \emph{traces} of classification outputs and policy decisions for a sample configuration of \fig{fig:output_correlation} ($sp$ restricts classification output metrics to a range of $10\%$ of the full set, while $rprob$ results in an average of $100$ images consecutively sampled from that range).  When compared to MDP, DQN recognizes when it enters periods of low metrics and proceeds to offload some low metric images while MDP does not.  Conversely, both policies perform mostly similarly during periods of high metric.

\begin{figure}[t]
	\centering
	\includegraphics[width=\linewidth]{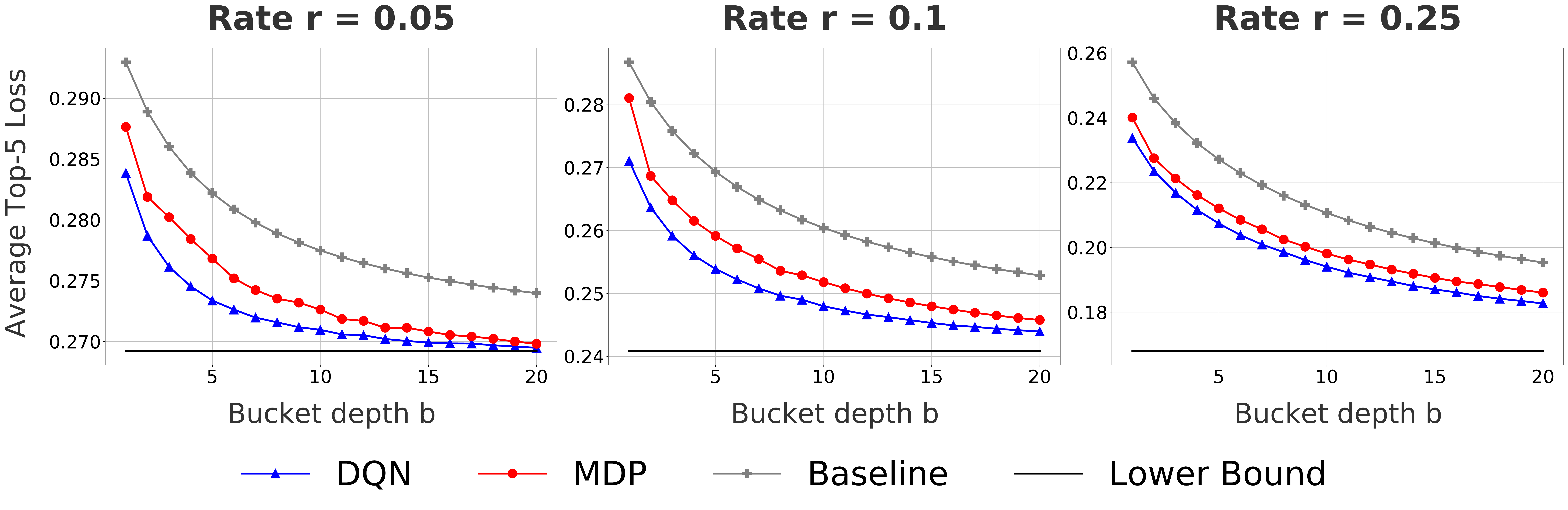}
	\caption{Offloading policies performance for different token bucket configurations under correlated classification outputs ($sp=0.1$ and $rprob=0.1$).
	}
	\label{fig:output_depth}
\end{figure}

\fig{fig:output_correlation} relied on a single token bucket configuration, $(r,b)=(0.1,4)$.  \fig{fig:output_depth} extends this by 
still relying on a particular pattern of classification output correlation ($sp=0.1$ and $rprob=0.1$), but now for different token bucket configurations.  Specifically, we select three different token rates, $r=0.05, 0.1, 0.25$ and for each vary the token bucket depth $b$ from 1 to 20.  The figure demonstrates that DQN consistently outperforms MDP and Baseline, even if the difference diminishes as either $r$ or $b$ increases.  This is expected.  A larger token rate lowers the cost of missed offloading opportunities because of wasted tokens, while a larger bucket depth makes offloading decisions less dependent on accurately predicting classification output correlation in successive images.  

We illustrate the latter in \fig{fig:output_depth_trace}, where we again plot traces of the decisions that the DQN and MDP policies make for a scenario with correlated output metrics ($sp=0.1$ and $rprob=0.01$) and two different bucket depths, $b=4$ (Top) and $b=20$ (Bottom).  We note that the value $rprob=0.01$ differs from that used in \fig{fig:output_depth}, \ie $rprob=0.1$.  The motivation is visual clarity, as the lower $rprob$ value stretches the periods during which classification output metrics are sampled from a given range, which amplifies differences in policy decisions.  Comparing the Top and Bottom parts of the figures, we see that when $b$ is larger, DQN recognizes that the odds of wasting tokens during periods of low metrics are lower, which results in fewer offloading decisions during those times. This is especially so after periods of high metrics, \eg after $t\approx 500$, when the token bucket count is low.

\subsubsection{Markov-modulated Image Arrivals and I.I.D.~Classification Outputs}
\label{sec:mma+iid}

Next, we proceed to demonstrate that DQN can also learn variations in the structure of the image \emph{arrival process,} and in particular changes in the arrival rate that extend over long enough periods of time to affect offloading decisions. As the focus is on variations in the image arrival process, we rely on a simple i.i.d.~structure for the classifier outputs.

\begin{figure}[t]
	\centering
	\includegraphics[width=\linewidth]{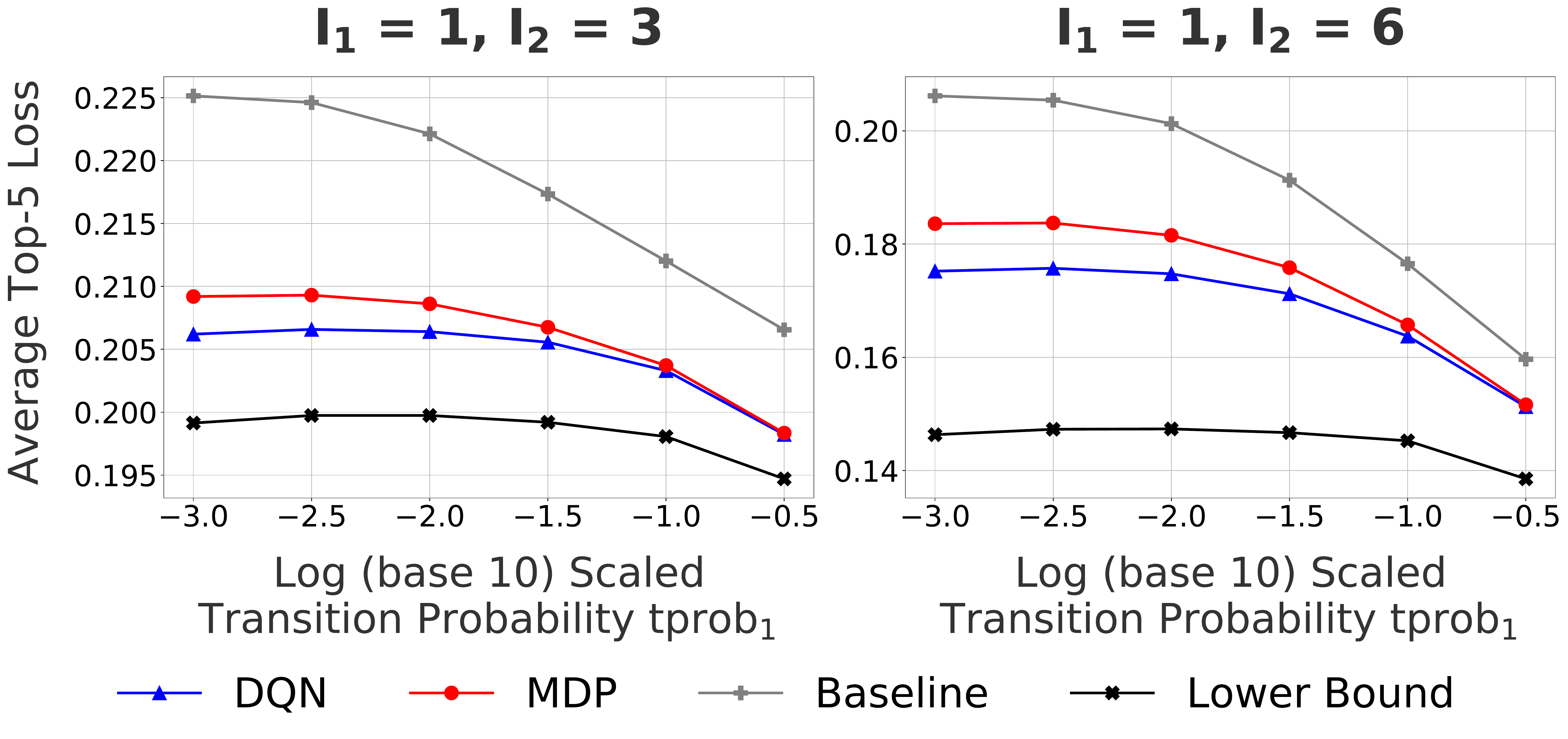}
	\caption{Offloading policies performance as function of image arrivals correlation.  Correlation decreases as transition probabilities $tprob_1$, $tprob_2$ increase. Token bucket: $r=0.1, b=4$.}
	\label{fig:input_correlation}
\end{figure}

As in the previous section, we chose $r=0.1, b=4$, as our base token bucket configuration, and evaluate offloading performance under Markov-modulated image arrival processes.  We rely on two base configurations.  Configuration~$1$ alternates between high and low intensity states with constant image inter-arrival times of $I_1=1$ and $I_2=3$, \ie images in every time slot versus every three time slots. We set the ratio of the transition probabilities out of each state to two, \ie $tprob_1/tprob_2=2$ so that the low intensity state lasts twice as long, and vary the state transition probability out of state~$1$, $tprob_1$, from $10^{-3}$ to $10^{-0.5}$.  Configuration~$2$ uses $I_1=1$ and $I_2=6$, \ie images again in every time slot in the high intensity state, but only every six time slots in the low intensity state, with $tprob_1/tprob_2=4$, \ie a low intensity state that now lasts four times as long.  As with the first configuration, we vary $tprob_1$ from $10^{-3}$ to $10^{-0.5}$.

The results are in \fig{fig:input_correlation}, which reports the average top-5 loss for DQN and our three benchmarks for configurations~$1$ (Left) and~$2$ (Right).  DQN's ability to learn the structure of the arrival process improves performance (lower Top-5 loss) over both MDP and Baseline, with those improvements diminishing\footnote{As mentioned in Section~\ref{sec:arrival}, changes in $tprob_i, i=1,2$ affect the image arrival rate. Hence, the changes in the lower bound as they increase.} as correlation in the arrival process decreases (increased transition probabilities out out each state).

\begin{figure}[t]
	\centering
	\includegraphics[width=\linewidth]{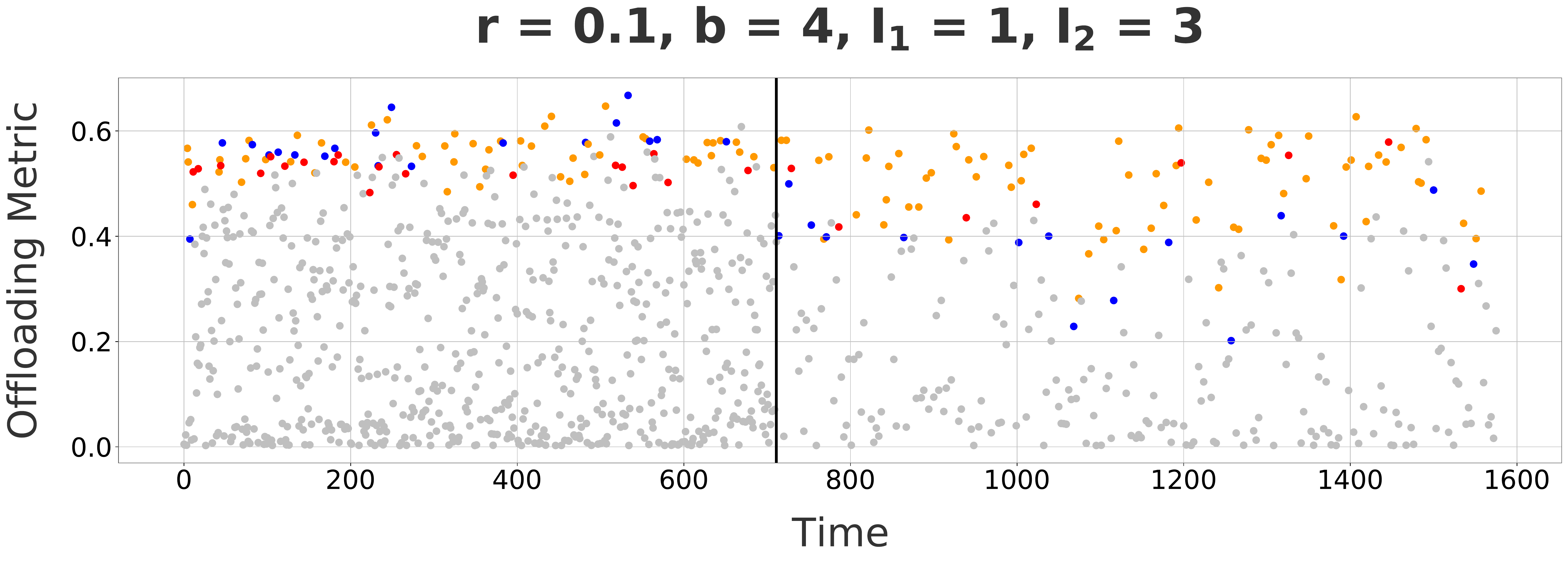}
	\includegraphics[width=\linewidth]{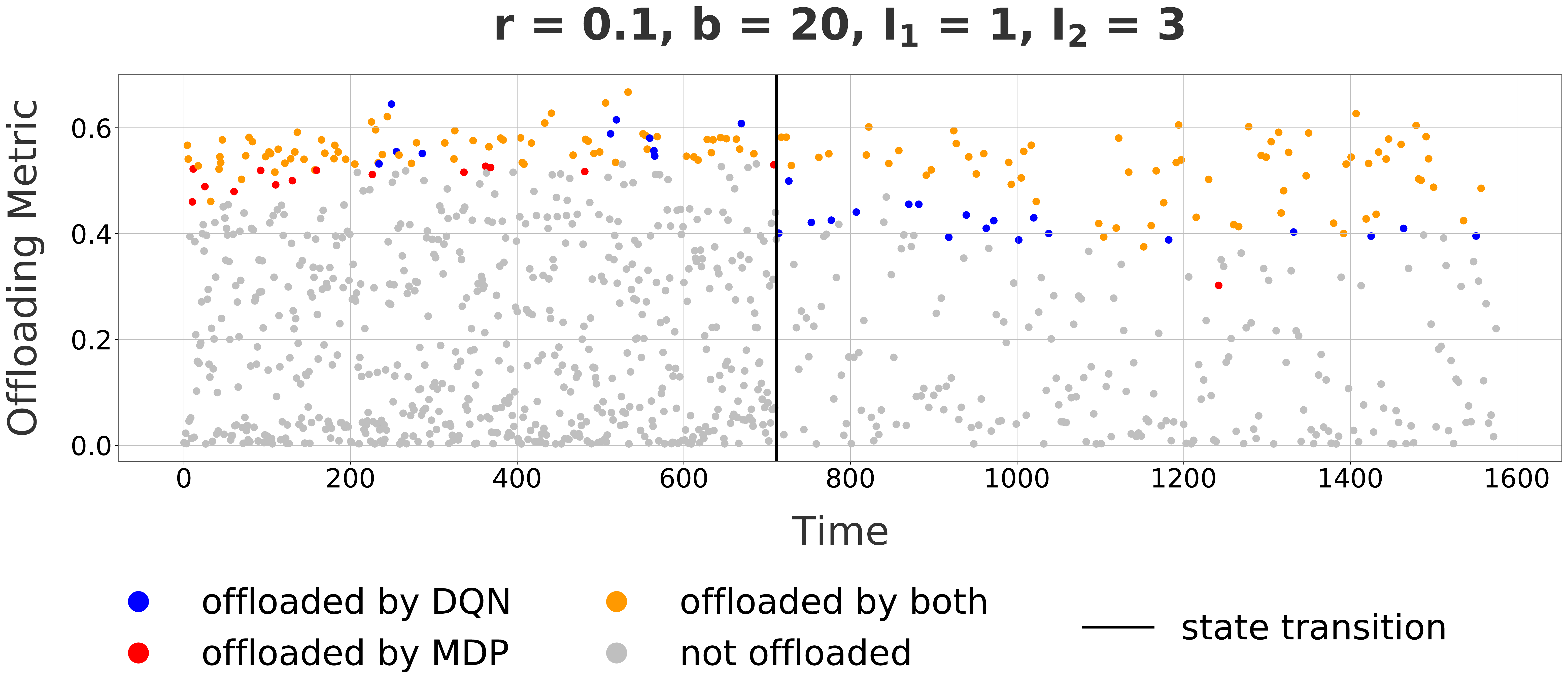}
	\caption{Offloading decisions of DQN and MDP for token bucket depths of $b=4$ (Top) and $b=20$ (Bottom) under correlated image arrivals ($I_1=1$, $I_2=3$, $tprob_1=0.001$, $tprob_2=0.0005$).}
	\label{fig:input_depth_trace}
\end{figure}

To better understand how learning about the arrival process affects DQN's offloading decisions, we again use a sample trace showing the decisions of both DQN and MDP for a sequence of image arrivals. To illustrate DQN's ability to ``recognize'' rate transitions, the trace explicitly includes one. The results are reported in the top portion of \fig{fig:input_depth_trace} for configuration~$1$ with state transition probabilities of $tprob_1=0.001$, $tprob_2=0.0005$.  The transition from high to low arrival intensity is indicated by a vertical line in the figure.  

In the high arrival rate state (left of the dividing line), DQN is more conservative than MDP with slightly fewer offloading decisions. This is, however, offset by its ability to offload some higher metric images than MDP whose more aggressive behavior resulted in an empty token bucket when those images arrived. Conversely, once DQN recognizes that it has transitioned to a state with a lower image arrival rate (right of the dividing line), it proceeds to be more aggressive and selects more lower metric images as it knows that the lower image arrival rate means that tokens will be replenished faster relative to image arrivals.  In contrast, MDP ends-up wasting tokens it could have used during periods of lower arrival rate.

\begin{figure}[t]
	\centering
	\includegraphics[width=\linewidth]{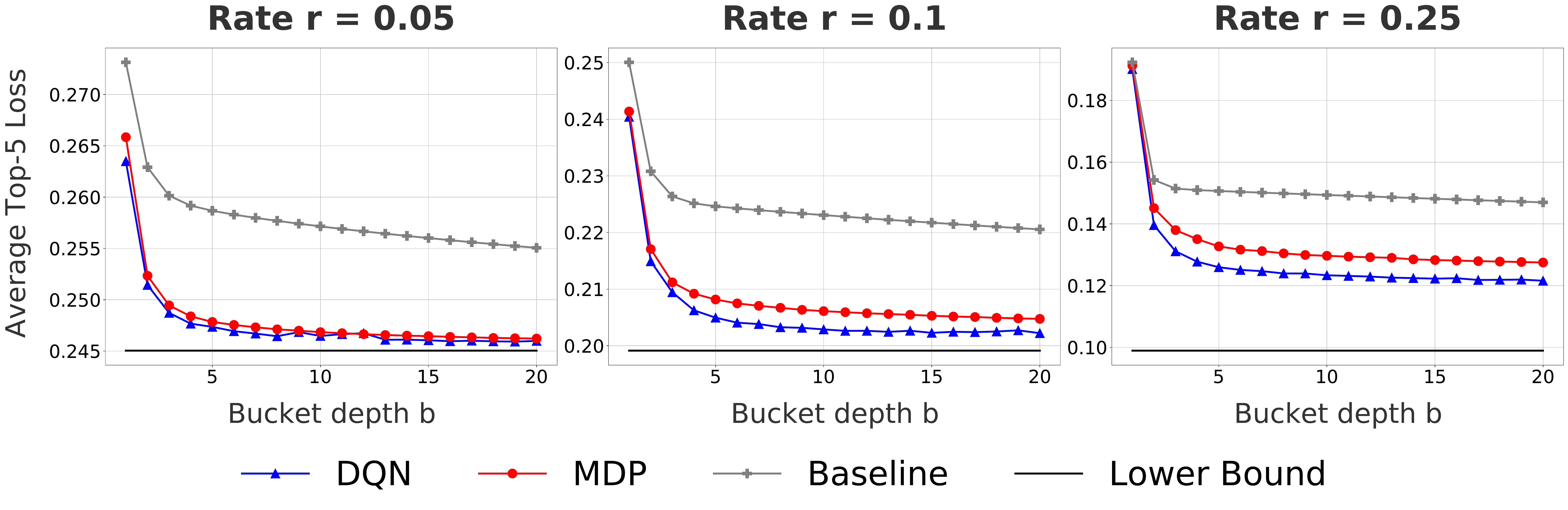}
	\caption{Offloading policies performance for different token bucket configurations under correlated image arrivals ($I_1=1$, $I_2=3$ and $tprob_1=0.001$, $tprob_2=0.0005$).}
	\label{fig:input_depth}
\end{figure}

Next, we investigate the extent to which the results of \fig{fig:input_correlation} remain under different token bucket configurations.  For that purpose, we select configuration~$1$ with $I_1=1, I_2=3$ and $tprob_1=0.001, tprob_2=0.0005$. \fig{fig:input_depth} reports the performance (top-5 loss) of DQN and our three benchmarks across a range of token bucket configurations, namely, token rates of $r=0.05, 0.1, 0.25$, and token bucket depths that vary from $b=1$ to~$20$.  The figure illustrates that DQN continues to outperform MDP across all configurations, even if, as with \fig{fig:output_depth}, the differences are smaller than between MDP and the Baseline.  The latter ignores the token bucket state, which remains the main contributor to differences in performance.

Towards better understanding factors that influence differences between DQN and the MDP policy in the presence of arrival correlation, the bottom part of \fig{fig:input_depth_trace} reports a trace of image arrivals ($I_1=1, I_2=3, tprob_1=0.001, tprob_2=0.0005$) and policy decisions that parallels that of the top part of the figure, but for a different token bucket depth, \ie $b=20$ versus $b=4$. The bigger bucket depth means that MDP's overly aggressive behavior during periods of high arrival rate (it still assumes the lower long-term rate) has less of an impact, as the larger bucket makes it easier to sustain the higher offloading rate (at least for a period of time).  This is illustrated by the fewer policy decision differences between MDP and DQN in the bottom part of the figure's left-hand-side. Conversely, the larger bucket also means that DQN needs not be as aggressive during periods of lower arrival rate since the larger bucket reduces the odds of wasting tokens by not offloading enough images. This is reflected in the higher metrics used by DQN in its offloading decisions in the right-hand-side of the bottom part of \fig{fig:input_depth_trace}.

\subsubsection{Markov-modulated Image Arrival and Correlated Classification Outputs}



\begin{figure*}[t]
	\centering
	\includegraphics[width=0.7\linewidth]{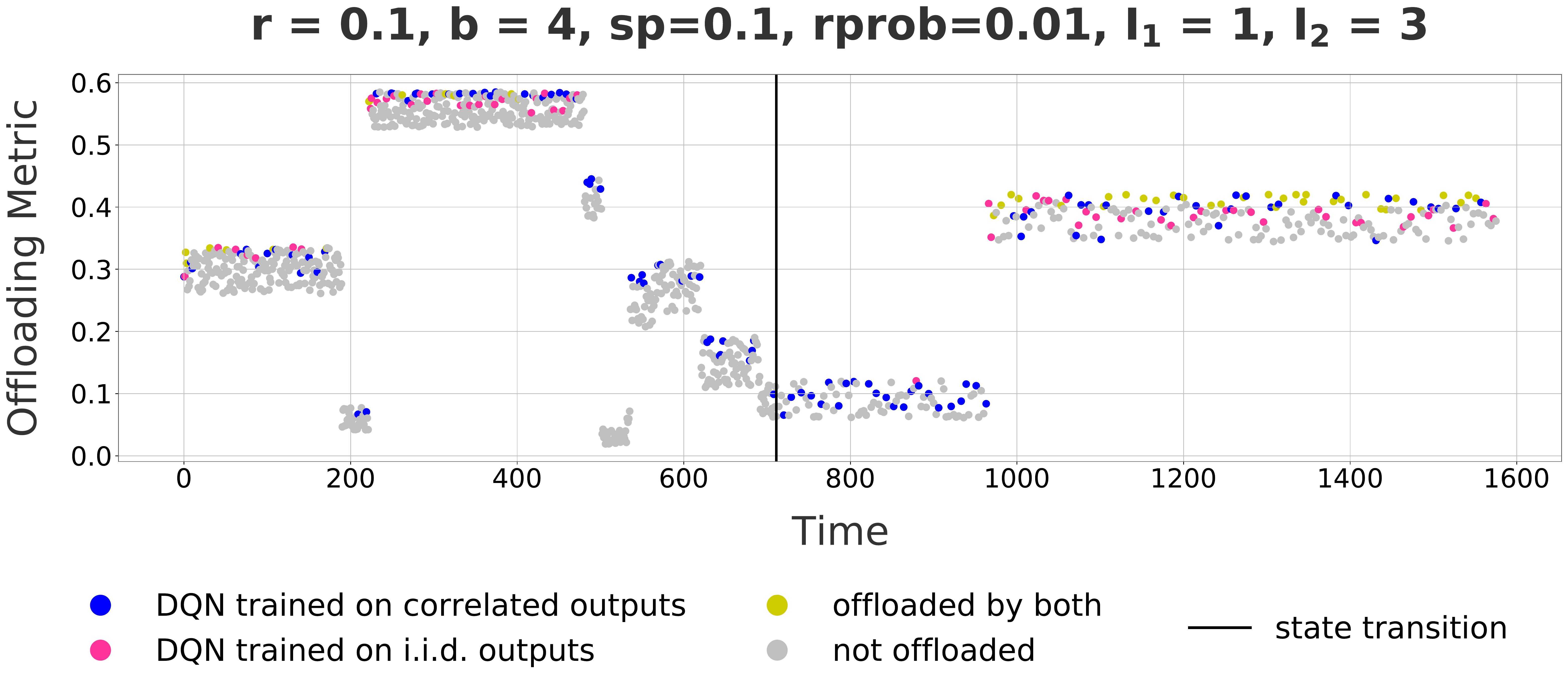}
	\caption{Offloading decisions of DQN (trained on the combined correlation) and DQN (trained on correlated arrivals with \iid classifier outputs) on a combined correlation setting.  For the token bucket configuration of $r=0.1$, $b=4$, we consider the image arrival $I_1=1$, $I_2=3$, $tprob_1=0.001$, $tprob2=0.0005$, combined with correlated classifier output of $sp=0.1$, $rprob=0.01$.  We compare the offloading decisions of DQN trained on this combined correlation setting, and DQN trained on the given arrival process, but \iid classifier outputs.}
	\label{fig:combine_trace}
\end{figure*}

This sub-section demonstrates DQN's ability to extract information about the structure of \emph{both} image arrivals and classification outputs, and to use it in its offloading decisions.  For that purpose, we combine the correlated image sequence generator of Section~\ref{sec:det+corr} with the Markov-modulated input process of Section~\ref{sec:mma+iid} to produce a sequence of arrivals with both variable arrival rate and correlation in the classification accuracy of successive outputs.

In keeping with the structure of Sections~\ref{sec:det+corr} and~\ref{sec:mma+iid}, we carried out an extensive set of experiments where we varied the image arrival process, classification output correlation, and token bucket parameters.  The goal was to offer a broad coverage of different configurations and evaluate DQN's performance across them. The results of those experiments were essentially similar to those found in earlier sections, with DQN outperforming the MDP and Baseline benchmarks across all configurations with differences of similar magnitude.  Because those results do not offer much additional insight, we omit them and instead focus on the analysis of a trace that helps shed some light on how DQN translates what it learns about the structure of its inputs into policy decisions.

This trace is reported in \fig{fig:combine_trace}.  It consists of image arrivals generated according to a Markov-modulated process similar to that of \fig{fig:input_depth_trace}, namely, $I_1=1$, $I_2=3$, $tprob_1=0.001$, and $tprob2=0.0005$, with classification outputs that exhibit the same correlation structure as in \fig{fig:output_depth_trace}, namely, $sp=0.1$ and $rprob=0.01$. In other words, the trace captures a sequence of inputs with both correlated arrivals and classification outputs.  As in most previous experiments, token bucket parameters were set to $r=0.1$ and $b=4$.

The figure reports (using again dots of different colors) the offloading decisions of two versions of the DQN policy.  Both were trained using the same sequences of image arrivals, but their training sequences differed in the structure of the corresponding classification outputs. The first version, $\text{DQN}^-$, was trained with i.i.d.~classification outputs, while the second, $\text{DQN}^+$, was trained with the correlated classification outputs present in the trace of \fig{fig:combine_trace} (the trace involves one transition in the arrival rate at $t = 711$ and multiple periods that span different ranges of offloading metrics on each side of that transition).  The purpose is to illustrate how learning about classification output correlation affects DQN's decisions given that it already knows about arrival correlation.  In other words, Sections~\ref{sec:det+corr} and~\ref{sec:mma+iid} demonstrated DQN's ability to learn and use structure in either the image arrival process or the classification output.  The intent is to show it can learn both by comparing two DQN versions that differ in the information available to them regarding classification outputs.

As in \fig{fig:input_depth_trace}, both policies are aware of possible changes in image arrival rates. Hence, they each offload more aggressively (lower metrics) after detecting a transition to a state with lower image arrival rate (a decrease of $20.27\%$ and $19.34\%$ for $\text{DQN}^+$ and $\text{DQN}^-$, respectively, in the metrics of offloaded images between before and after $t = 711$). The more interesting aspect though is in the differences in decisions between the two policies as correlation in classification outputs produces periods with different correlation ranges, a phenomenon that was not incorporated in the training of $\text{DQN}^-$.  This is best seen by focusing on two specific regions for which this is more visually apparent, namely, a period of relatively high metrics in the interval $t\in[222,480]$ and conversely a period of relatively low metrics in the interval $t\in[691,963]$.

During the high metrics interval, $\text{DQN}^+$ becomes aware that it can expect consistently higher metrics, and so adjusts its decisions to be more conservative.  Both policies offload roughly the same number of images, $27$ and $29$ for $\text{DQN}^+$ and $\text{DQN}^-$, respectively, but the average metric of images offloaded by $\text{DQN}^+$ is $0.580$ versus $0.572$ for $\text{DQN}^-$, a small but meaningful difference, especially given the narrow range of metrics sampled during that period (between $0.529$ and $0.585$). Conversely, during the low metrics interval, $\text{DQN}^+$ realizes that it will be getting images with relatively low metrics for some time, and consequently lowers its expectations and offloads lower metric images to avoid wasting tokens.  This results in $\text{DQN}^+$ offloading $24$ images during that period versus only one image for $\text{DQN}^-$.  Those differences highlight how the $\text{DQN}^+$ policy leverage the additional information it learned about the structure of classification outputs.  In turn, those resulted in improved performance with average top-5 losses of $0.247$ and $0.253$ for $\text{DQN}^+$ and $\text{DQN}^-$, respectively.


\subsubsection{DQN Modeling Parameters}
\label{sec:dqn_param}

\begin{figure}[t]
	\centering
	\includegraphics[width=\linewidth]{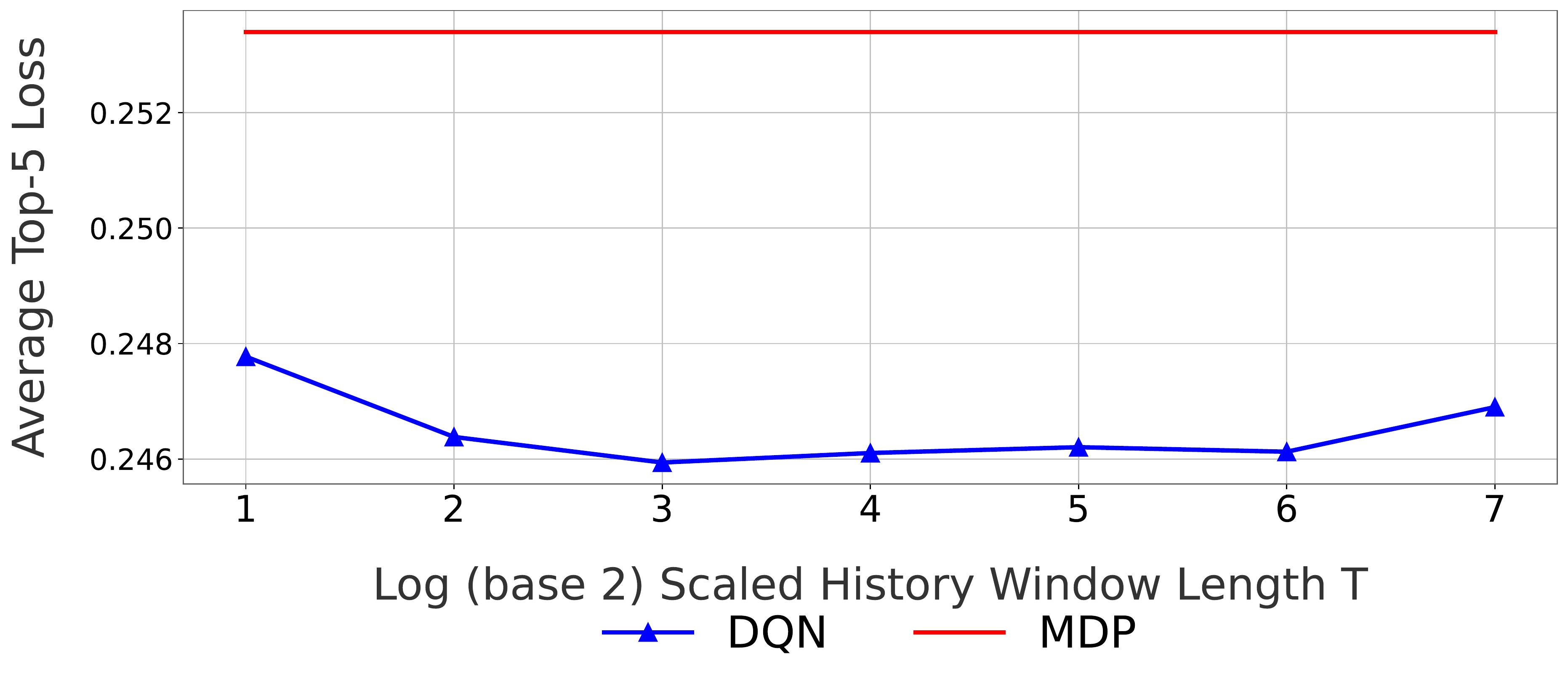}
	\caption{DQN's performance as a function of the log (base 2) of the history window length $T$ on a representative setting that combines correlation in both image arrivals ($I_1=1$, $I_2=3$, $tprob_1=0.001$, $tprob2=0.0005$) and classification outputs ($sp=0.1$, $rprob=0.01$), and token bucket parameters of $r=0.1$ and $b=4$.}
	\label{fig:T}
\end{figure}

In this last sub-section we investigate how the 
DQN's parameters, including the history window length $T$, the number of layers, and the number of units in each layer,
impact performance of our policy.  We report results for a setting that combines both variable image arrival rate and correlated classification output, as it represents a more complex environment for which the choice of window length can, therefore, be anticipated to have a greater impact.

\fig{fig:T} reports the performance (average top-5 loss) of DQN (and MDP\footnote{MDP is included only to show that DQN outperforms it for all $T$ values.}) for different values of $\log_2T$. The lowest value $(T=2)$ corresponds to a setting where DQN uses only the current offloading metric and image inter-arrival time, while the largest setting of $T=128$ offers enough samples for DQN to learn the correlation structure in both arrivals and classification outputs.

The results display relatively limited sensitivity to the choice of $T$ even if some variations are present.  Of note is the fact that even in the absence of any history $(T=2)$, DQN still outperforms MDP because it can use its knowledge of the current inter-arrival  time to  make better policy decisions (MDP only has access to the current offloading metric and token bucket state).  As $T$ increases and more history information becomes available, DQN quickly stabilizes at its best performance and remains insensitive to $T$ over a wide range.  Performance eventually starts to decrease as $T$ becomes too large. This is likely because its simple architecture (a $5 \times 64$ MLP) does not contain a sufficiently large number of parameters to interpret all the information within the high-dimensionality input.

We also performed a grid search on the model parameters by varying the number of hidden layers from $3$ to $8$, and the base $2$ logarithm of the number of units in each layer from $4$ to $8$.  As when varying the history window length $T$, we observed only small variations (within $1\%$) in the relative difference between the best and the worst performance for the top-5 loss. This indicates limited sensitivity of the model to these choices.

\section{Conclusion}

The paper investigates a distributed image classification problem in an edge-assisted AIoT setting, where classification accuracy is improved by dynamically offloading some images to an edge server subject to network bandwidth constraints.  Managing access to the shared network is regulated through a token bucket that constrains offloading decisions.
The paper devises and evaluates a policy that manages offload decisions from devices under such constraints while optimizing classification accuracy. Because image arrival patterns and classification results can be arbitrary, the policy needs to accommodate complex input sequences.  To that end, we investigate the use of DQN to realize such a policy, and demonstrate its ability to effectively ``learn'' 
effective policy decisions.  Experiments demonstrate both the efficacy of the DQN-based offloading policy and its runtime efficiency on embedded devices with limited computational resources.

\bibliographystyle{IEEEtran}
\bibliography{reference}
\clearpage

\end{document}